\documentclass[usenatbib]{mnras}

\usepackage{newtxtext,newtxmath}
\usepackage[T1]{fontenc}
\usepackage[utf8]{inputenc}
\usepackage{ae,aecompl}

\usepackage{graphicx}	
\usepackage{amsmath}	
\usepackage{amssymb}	

\newcommand{\pc}{\,\mathrm{ pc }}
\newcommand{\Mdot}{\,\mathrm{ M}_{\odot} }
\newcommand{\kpc}{\,\mathrm{ kpc }}
\newcommand{\Myr}{\,\mathrm{ Myr }}

\newcommand{\km}{\,\mathrm{ km }}
\newcommand{\cm}{\,\mathrm{ cm }}
\newcommand{\s}{\,\mathrm{ s }}
\newcommand{\G}{\mathrm{G}}

\renewcommand{\d}{\mathrm{d}}

\title[The birth of a supermassive black hole binary]{The birth of a supermassive black hole binary}

\author[H. Pfister et al.]{
Hugo Pfister,$^{1}$\thanks{E-mail: pfister@iap.fr}
Alessandro Lupi,$^{1}$
Pedro~R. Capelo,$^{2}$
Marta Volonteri,$^{1}$
\newauthor{
Jillian~M. Bellovary$^{3}$
and Massimo Dotti$^{4,5}$}
\\
$^{1}$Sorbonne Universit\'{e}s, UPMC Universit\'{e} Paris 06 et CNRS, UMR7095, Institut d'Astrophysique de Paris,\\
98bis boulevard Arago, F-75014, Paris, France\\
$^{2}$Center for Theoretical Astrophysics and Cosmology, Institute for Computational Science, University of Zurich,\\
Winterthurerstrasse 190, CH-8057 Z$\ddot{u}$rich, Switzerland\\
$^{3}$Department of Physics, Queensborough Community College, 222-05 56th Avenue, Bayside, NY 11364, USA\\
$^{4}$Dipartimento di Fisica G. Occhialini, Universit$\grave{a}$ degli Studi di Milano--Bicocca, Piazza della Scienza 3, I-20126 Milano, Italy\\
$^{5}$INFN, Sezione Milano--Bicocca, Piazza della Scienza 3, I-20126 Milano, Italy
}

\date{Accepted XXX. Received YYY; in original form ZZZ}

\pubyear{2017}

\begin{document}
\label{firstpage}
\pagerange{\pageref{firstpage}--\pageref{lastpage}}
\maketitle

\begin{abstract}
We study the dynamical evolution of supermassive black holes, in the late stage of galaxy mergers, from kpc to pc scales. In particular, we capture the formation of the binary, a necessary step before the final coalescence, and trace back the main processes causing the decay of the orbit. We use hydrodynamical simulations of galaxy mergers with different resolutions, from 20~pc down to 1~pc, in order to study the effects of the resolution on our results, remove numerical effects, and assess that resolving the influence radius of the orbiting black hole is a minimum condition to fully capture the formation of the binary. Our simulations include the relevant physical processes, namely star formation, supernova feedback, accretion onto the black holes and the ensuing feedback. We find that, in these mergers, dynamical friction from the smooth stellar component of the nucleus is the main process that drives black holes from kpc to pc scales. Gas does not play a crucial role and even clumps do not induce scattering or perturb the orbits. We compare the time needed for the formation of the binary to analytical predictions and suggest how to apply such analytical formalism to obtain estimates of binary formation times in lower resolution simulations.  
\end{abstract}



\section{Introduction}
\label{Introduction}
Observations indicate that most massive galaxies host in their centres a supermassive black hole (SMBH) weighing millions to billions of solar masses \citep{Kormendy_95}. Those SMBHs are thought to co-evolve with their host galaxy, as suggested by relations  between the mass of SMBHs and their host galaxy properties \citep[e.g. bulge mass, velocity dispersion, for a review see][and references therein]{KormendyHo_13}.

This shows how crucial it is to study the evolution of SMBH mass, which has two ways of growing: either via accretion of gas and stars, or via mergers with other SMBHs \citep{Volonteri_03, Dubois_14, Sesana_14}. The latter,  SMBH mergers, happen when two galaxies, hosting in their centre a  SMBH, collide and merge.

On large scales, dynamical friction is the main process that brings the SMBHs closer \citep{Begelman_80,Yu2002,Callegari_11, Callegari_09, Capelo_15}. However, dynamical friction becomes inefficient when the two SMBHs form a bound binary, and they are close enough that their orbital velocity becomes larger than the velocity dispersion of the surrounding stars \citep[e.g.][]{Begelman_80,Quinlan1996,Milosavljevic2001,2006ApJ...642L..21B,2015ApJ...810...49V}, at $\sim$ pc scale. The subsequent evolution is driven by different mechanisms, e.g. scattering with individual stars \citep{Begelman_80} or viscous drag in a circumbinary disc \citep{Cuadra_09,Roedig_11,DelValle_12}. Ultimately, when the two SMBHs are close enough, at separations of order of mpc, they merge by emitting gravitational waves \citep{Begelman_80,Wyithe_03,Sesana_04,Sesana_05}.

In this paper, we focus on the formation of a bound  SMBH binary (SMBHB). Typical simulations of galaxy mergers do not have the resolution to capture the formation of a SMBHB, which occurs on pc scales. For this reason we zoom in on the high spatial (20~pc) and temporal (1~Myr) resolution simulations of galaxy mergers by \cite{Capelo_15} to capture the formation of the binary. 

\section{Numerical set up}
\label{Numerical_set_up}
In this Section, we briefly describe the main characteristics of the original simulations. We then present the new runs we performed for our study. We used the $N$-body smoothed particle hydrodynamics code {\scshape gasoline} \citep{Wadsley_04}, an extension of the pure gravity tree code {\scshape pkdgrav}  \citep{Stadel_01}. The version we used includes explicit line cooling for atomic hydrogen and helium, and metals \citep{Shen_10}, a physically motivated prescription for star formation (SF), supernova feedback, and stellar winds \citep{Stinson_06}, as well as SMBH accretion and feedback \citep{Bellovary_10}.

\subsection{Original simulation}
\label{Initial_simulation}
\begin{table*}
 \begin{tabular}{lccccccp{7.5cm}}
  \hline
	Name&
	$\epsilon_\text{gas}$&
	$\epsilon_\text{star}$&
	$\epsilon_\text{BH}$&
	$\rho_{\text{crit}}$&
	$\tau_\text{SMBHB}$&
	$\tau_\text{res}$&
	Description\\
	
	&
	pc&
	pc&
	pc&
	100 a.m.u. cm$^{-3}$&
	Myr&	
	Myr
	\\
	
	\hline
	\hline
	
	R20 & $20$ & 10 & 5 & $1$&55 & 24 &Same resolution as the original run but ``trimming'' the galaxy\\
	R5 & $5$ & 2.5 & 1.25 &$10$&27 & 14 &X\\
	R2 & $2$ & 1 & 0.5 &$60$&19 & 18 &X\\
	R1 & 1 & 0.5 & 0.25 &240 &19 & 19 &X\\
	
	\hline
	
	R2b & 2  &  1 & 0.5 & 60 &19 & 19 &Begins at 12 Myr, BH1 shifted by 3 pc\\
	R2c & 2  & 1 & 0.5 & 60 &20 & 19 &Begins at 12 Myr, BH2 velocity increased by 20 per cent\\
	R2d & 2  & 1 & 0.5 & 60 &19 & 19 &Begins at 12 Myr, BH1 shifted by 16 pc\\
		
	\hline
  	
  	R5\_1to2 & $5$ & 2.5 & 1.25 & $10$& 43 & 26 & 1:2 mass ratio, no nuclear coup\\	
	R5\_Inclined & $5$ & 2.5 & 1.25 & $10$ & 348 & 373 &Inclined orbit, no nuclear coup\\  	  	
  	\hline
 \end{tabular}
 \caption{Simulations performed. We vary the softening length, $\epsilon$, and the density threshold for SF, $\rho_{\text{crit}}$. $\tau_\text{SMBHB}$ is the time at which the SMBHB is formed in our simulations and $\tau_\text{res}$ corresponds to the moment the distance between SMBHs is below $\epsilon_{\rm gas}$ for the first time. A description of the different simulations is also given.}
 \label{tab_simulation}
\end{table*}
Among all the simulations presented in \cite{Capelo_15}, we first zoom in on the 1:4 coplanar, prograde--prograde merger of galaxies (namely Run 07 in \citealt{Capelo_15}, hereafter the ``original simulation'').

At the beginning of the original simulation, there are two coplanar galaxies, one (G1) being four times more massive than the other (G2), both hosting in their centres a SMBH (BH1 and BH2), whose masses are proportional to the mass of the bulge of each host galaxy. We refer to \cite{Capelo_15} for a detailed description of the initial set up.

We chose this particular simulation because the mass ratio 1:4 is usually chosen as the boundary between major and minor mergers. The merger time-scale in major mergers is shorter,  as the dynamical friction time-scale  $\propto 1/M_{\text{Satellite}}$ \citep{Chandrasekhar_43,BT_87}, where $M_{\text{Satellite}}$ is initially the mass of the lighter galaxy, then that of the stellar nucleus and, at the end, the mass of the orbiting SMBH. We expect therefore that forming a SMBHB is easier in mergers of similar-mass galaxies. For instance, in simulations with small mass ratios, down to 1:10 in \cite{Capelo_15}, the time needed for the two SMBHs to reach kpc separation (from an initial separation equal to the sum of the virial radii of the merging galaxies) is roughly 3~Gyr, much longer than the $\sim$1~Gyr needed in the 1:4 simulation. As a consequence, we expect binaries resulting from minor mergers to be rarer. However, major mergers are less common than minor mergers \citep{Fakhouri_10} and therefore do not comprise the bulk of the merging population. A mass ratio of 1:4 appears to be a reasonable compromise between the rarity of the  galaxy merger itself and the duration of the merger process.  Additionally, in this particular simulation, a nuclear coup occurs (see \citealt{VanWassenhove_14} for details on nuclear coups):  the nucleus of G1, N1, is completely disrupted by tidal forces and BH1, which is more massive than BH2, becomes a satellite and orbits around BH2 and N2, the nucleus of the secondary galaxy. Since, as noted above, the time needed for the decay driven by dynamical friction scales as $1/M_{\text{Satellite}}$, the orbital decay is faster when the orbiting SMBH is the most massive of the two. We also treat the case without a nuclear coup in Section~\ref{OtherMergers}.

We estimate here the time needed to form a SMBHB, starting from $t_0$, where $t_0$ corresponds  to the time of the snapshot, in the original simulation, closest in time to that of the first apocentre, in the merger phase \citep[see][]{Capelo_15}, when the distance between the two SMBHs is smaller than 1.2~kpc. In the original simulation, $t_0=$1.20~Gyr after the beginning of the merger \citep[see][]{Capelo_15}.
Our criterion to determine when the SMBHB is formed is the same as in \cite{VanWassenhove_14}: a binary forms when the SMBH separation remains below $a$, the radius at which the total enclosed mass, excluding SMBHs, is equal to twice the combined mass of the SMBHs: $M_\text{tot}(a)=2\left(M_\text{BH1}+M_\text{BH2}\right)$. For the parameters of our study, $a\lesssim10$~pc. This definition allows us to distinguish between the formation of the SMBHB and a pericentre where the two SMBHs are very close but with a high relative velocity that takes them to larger distances afterwards. In the initial simulation (see Section \ref{Zoom_in}), with this definition, the SMBHB is formed at  $\tau_\text{SMBHB}=55$~Myr. However, since the original simulation's resolution for gas (20~pc) and stars (10~pc) is larger than the typical SMBH separation needed to form a SMBHB ($\lesssim$10~pc), the SMBH dynamics cannot be followed very accurately in the final stages of the pairing. For instance, in the initial simulation, the distance between the two SMBHs is smaller than the gravitational resolution (20~pc) for the first time at $\tau_\text{res}=24\Myr<\tau_\text{SMBHB}=55\Myr$, thus we expect the dynamics not to be captured properly afterwards. In order to address this issue, we increase the spatial resolution in the nucleus in a new set of simulations, which are described in the next Section.

\subsection{Zoom-in simulations}
\label{Zoom_in}

We begin our zoom-in simulations at $t_0$, which we now denote by $t_0=0$, and evolve the system for 30~Myr in order to capture the formation of the SMBHB.

With the aim of reducing the computational time and increasing the resolution at no additional cost, we first removed 3 million particles over the 8.5 million that were in the original simulation. The removed particles have been selected as the particles outside a radius of 20 kpc from the system's centre, assuming that they cannot affect the dynamics of the central kpc, where the SMBHs are orbiting at $t_0$. As a test run, we performed a simulation (hereafter the ``initial simulation''; R20 in Table \ref{tab_simulation}) at the same resolution of the original simulation, but without these outer particles. We then compared different quantities (density, average radial speed, average tangential speed or the orbit of SMBHs, see Appendix \ref{sec:EffectsOfTrimmingTheGalaxy}) between the original and initial simulation and found that, even after 90 Myr, the difference was very small in the inner 10 kpc, confirming our expectations.

\begin{figure}
 \includegraphics[width=\columnwidth]{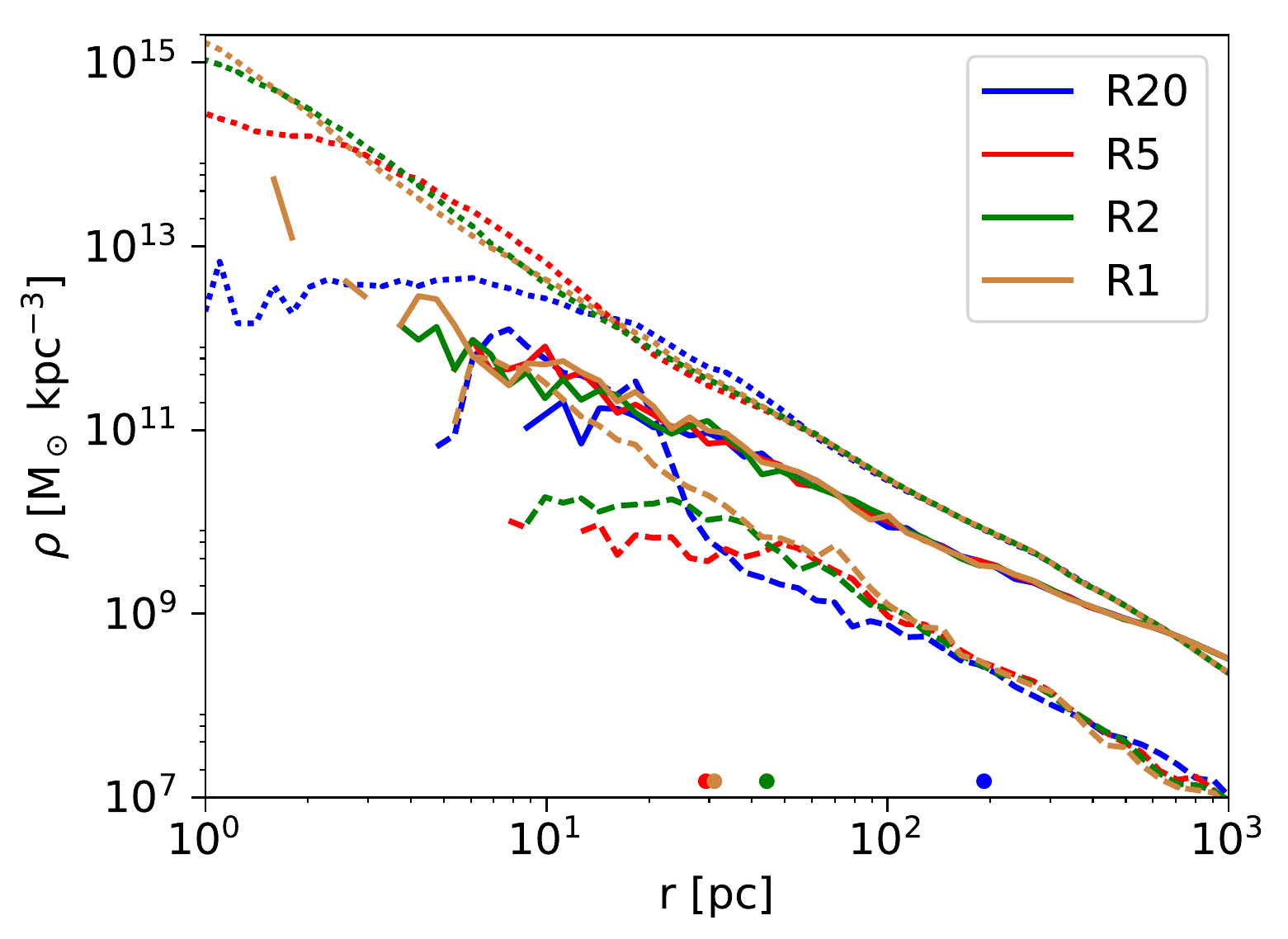}
 \caption{Density profile, centered on BH2, for gas (dashed), stars (dotted), and dark matter (solid) for different runs of the 1:4 coplanar, prograde--prograde merger, 17 Myr after $t_0$. We see that the profiles are fairly similar for all components except for stars in the inner 20 pc. Dots indicates the position of BH1.}
 \label{fig:DensityAllComponents}
\end{figure}

We increase the spatial resolution of the simulation by decreasing the value of the softening length, $\epsilon$, of all particles (gas, SMBHs, stars, and dark matter). This has the potential effect of making dark matter more collisional, while it should mostly represent a smooth potential. However, the dark matter density profile is not substantially affected by the increased resolution, as shown in Fig. \ref{fig:DensityAllComponents}. Moreover, as explained in \cite{Bellovary_10}, adopting dark matter particle masses that are similar to gas particle masses minimizes the effects of two-body interactions, which greatly helps to avoid spurious oscillations of SMBHs. This translates into a linear rescaling of all the softening lengths, if we assume that the density ratio between different components should remain constant. The overall effect is to increase the global resolution of the simulation. Similarly, the gas density profile is not affected. However, for a gas particles of mass $M$, decreasing the softening length also allows lower values of the minimal smoothing length, being $h_{\text{min}}=0.1\epsilon$. In dense regions, this inevitably leads to higher density $\rho$, the relation between those quantities being:
\begin{eqnarray}
\rho\propto M h^{-3}. 
\label{eq:MassToRho}
\end{eqnarray}
SF is allowed following the recipe from \cite{Stinson_06}. In particular, a gas particle can form stars if
\begin{eqnarray}
\rho&\geq&\rho_{\text{crit}} \, ,\label{eq:rhorhocrit} 
\end{eqnarray}
where $\rho_{\text{crit}}$ is a free parameter. Since $\rho$ varies with resolution, $\rho_\text{crit}$ must be tuned when resolution is changed. We rescale $\rho_\text{crit}$ so that gas particles form stars when they are Jeans unstable at a fixed temperature. This means $M\geq\left( k_\text{B} T\G m_\text{p}\right)^{3/2}\rho^{-1/2}$, where $k_\text{B}$ is the Boltzmann constant, $m_\text{p}$ is the proton mass, $\G$ is the gravitational constant and $T$ is the temperature floor of the cooling function. Using Eq.~(\ref{eq:MassToRho}), we find $\rho \propto h^{-2}$. For this reason we chose $\rho_{\text{crit}}\propto h^{-2}\propto\epsilon^{-2}$, which gives us how to tune $\rho_{\text{crit}}$ as a function of $\epsilon$. We show the effects of this change in Appendix \ref{sec:EffectsOfZoomingInOnTheStarFormation}  and note that the stellar  density is in good agreement with the value of $10^{15} \Mdot \kpc ^{-3}$ found by \cite{Schodel_14} 1 pc away from Sagittarius A$^\star$ in the Milky Way.

We decided not to change the mass resolution, i.e. the mass of individual particles, for two reasons. First, a higher mass resolution, i.e. a smaller particle mass, also corresponds to a larger number of particles, which inevitably leads to higher computational costs. Moreover, in the original simulation, the mass resolution is already high (a few 10$^3\Mdot$), much smaller than the typical mass of SMBHs ($\gtrsim 10^6\Mdot$). Second, our SF recipe is based on the Kennicutt--Schmidt law (\citealt{Kennicutt_98}; \citealt{Schmidt_59}; \citealt{Stinson_06}), which describes well the average SF on large scales, e.g. galactic discs or molecular clouds. A reduction of the particle mass to less than $10^3 \Mdot$ would require a different prescription for both SF and supernova feedback, which is beyond the scope of the present study.

The runs performed are listed in Table \ref{tab_simulation}. There are three sets of simulations. In the first set, we simply increase the resolution of the initial simulation to capture the dynamics of the SMBHs. The resolution is progressively increased to be able to discriminate between numerical effects and new phenomena captured owing to the higher resolution. We use the second set to determine if the trajectory of the SMBHs depends on our initial parameters. We do this by re-simulating the R2 simulation ($\epsilon_{\rm gas} = 2$~pc), 12~Myr after the beginning of R2, but shifting the position of BH1 (keeping the distance between the SMBHs constant) or increasing the velocity of BH2. Finally, we perform two other runs, zooming in on the 1:2 coplanar, prograde--prograde simulation and on the 1:4 inclined-primary simulation (Runs 02 and 08 in \citealt{Capelo_15}), where no nuclear coup occurs, to investigate how the initial orbital inclination, mass ratio of the two galaxies, and the presence/absence of a nuclear coup impact our results. We adopt the same technique used for the 1:4 coplanar, prograde--prograde merger to perform those simulations, removing particles that are farther than 20~kpc from the centre of the system and increasing the spatial resolution by a factor of four, reaching $\epsilon_{\rm gas} = 5$~pc.

\section{Dynamical evolution}
\label{DynamicalEvolutionOfBHs}
In Section~\ref{AFasterDecay}, we study the orbital evolution of the SMBHs in the different simulations of the 1:4 coplanar, prograde--prograde merger. In Section~\ref{Explanation:DynamicalFriction}, we show how dynamical friction from stars drives the formation of the SMBHB.

\subsection{A faster decay}
\label{AFasterDecay}

\begin{figure}
 \includegraphics[width=\columnwidth]{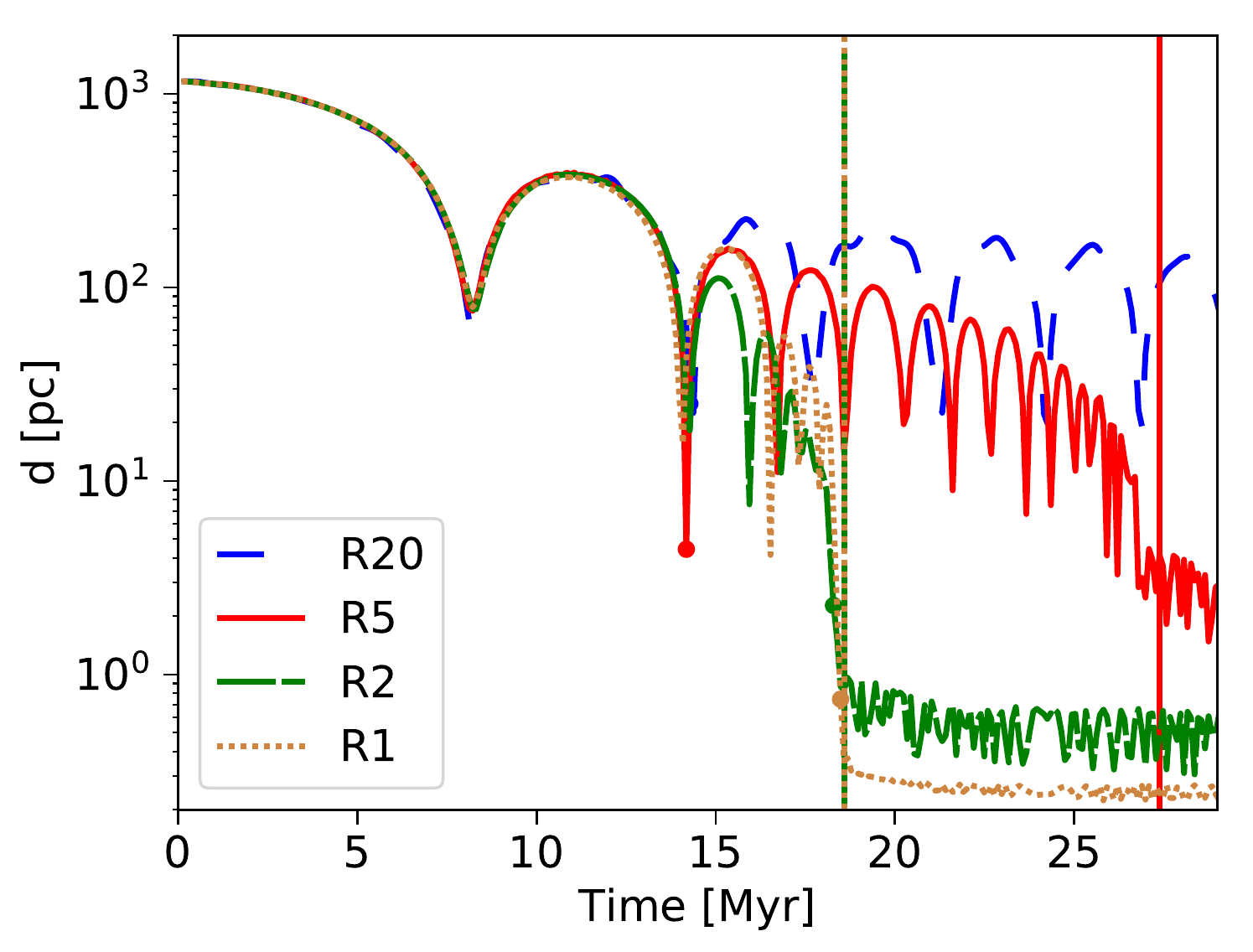}
 \caption{Distance between the two SMBHs for different runs of the 1:4 coplanar, prograde--prograde merger. The vertical lines show when the SMBHB is formed. The dots indicate the first time the distance between the two SMBHs is below resolution. In R20, the binary forms at 55 Myr. All quantities are shown as a function of time.}
 \label{fig:DistBHvsTime}
\end{figure}

In Fig.~\ref{fig:DistBHvsTime}, we show the distance between the two SMBHs as a function of time and the time at which the two SMBHs form a SMBHB, according to our criterion given in Section~\ref{Initial_simulation}. In runs with a resolution better than 20~pc (hereafter ``high-resolution runs''), we find that the binary is formed in about 20~Myr and confirm that the original simulation did not have the resolution to capture the final stages of the SMBH pairing.

We give here two possible explanations for the sharp decay of the distance we observe in R1 and R2 and discuss them in more detail in Sections~\ref{Explanation:DynamicalFriction} and \ref{EffectsOfGasClumps}:
\begin{itemize}
\item One possibility is that, due to the relatively large value of the gravitational softening in the initial simulation, the nucleus would not be sampled well and could be less dense than in reality. As a consequence, dynamical friction on BH1, which scales linearly with density \citep{BT_87}, would be less effective in the initial run, resulting in a longer pairing time-scale. Force resolution and gravity are also a key element in determining the evolution of the system as detailed in Section~\ref{Explanation:DynamicalFriction}.
\item The other possibility is that, in principle, an increased resolution allows us to better resolve clumps of material that were smoothed in the original and initial simulations. This would lead, if those clumps are massive enough, to SMBH-SMBH-Clump+background interactions instead of simple SMBH-SMBH+background interactions. Also, the trajectory and the orbit of SMBHs could be strongly affected \citep{Fiacconi_13,  Lupi_15, SouzaLima_17,Tamburello_17}. This is discussed in Section~\ref{EffectsOfGasClumps}.
\end{itemize}

\subsection{The role of dynamical friction}
\label{Explanation:DynamicalFriction}

\begin{figure}
 \includegraphics[width=\columnwidth]{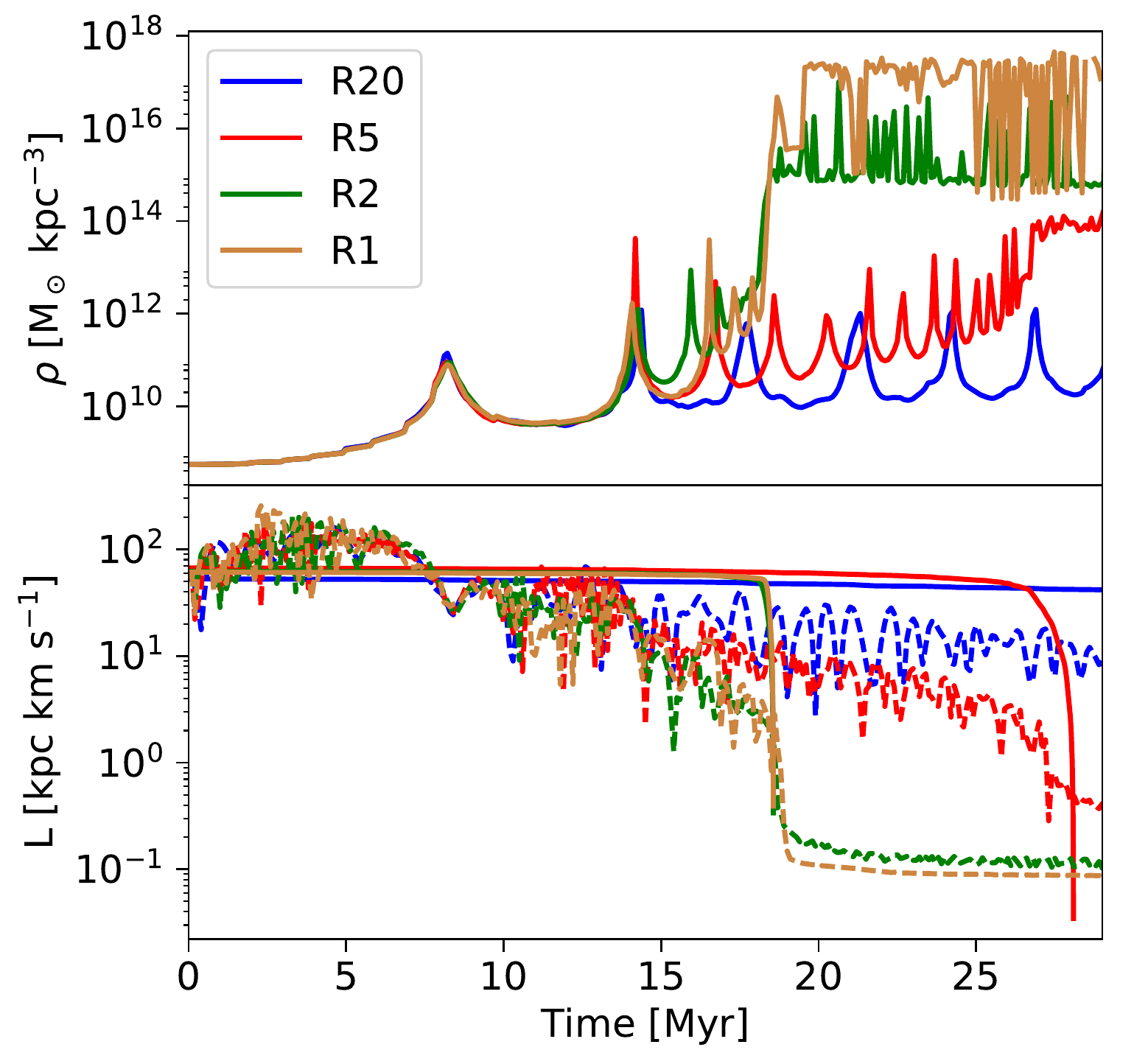}
 \caption{Top panel: density at the position of the orbiting SMBH, BH1, for all resolutions as a function of time. At $t=13$~Myr, the density seen by BH1 is the same in all simulations, but, while in R1 and R2 it is captured in the nucleus of BH2, in R5 and R20 the SMBH escapes. Bottom panel: result of the integration of Eq.~(\ref{eq:DynamicalFriction}) (solid) and specific angular momentum measured in the simulations (dashed) as a function of time.}
 \label{Fig:RhoAndV}
\end{figure}

\begin{figure*}
\centering
\includegraphics[height=0.22\textheight]{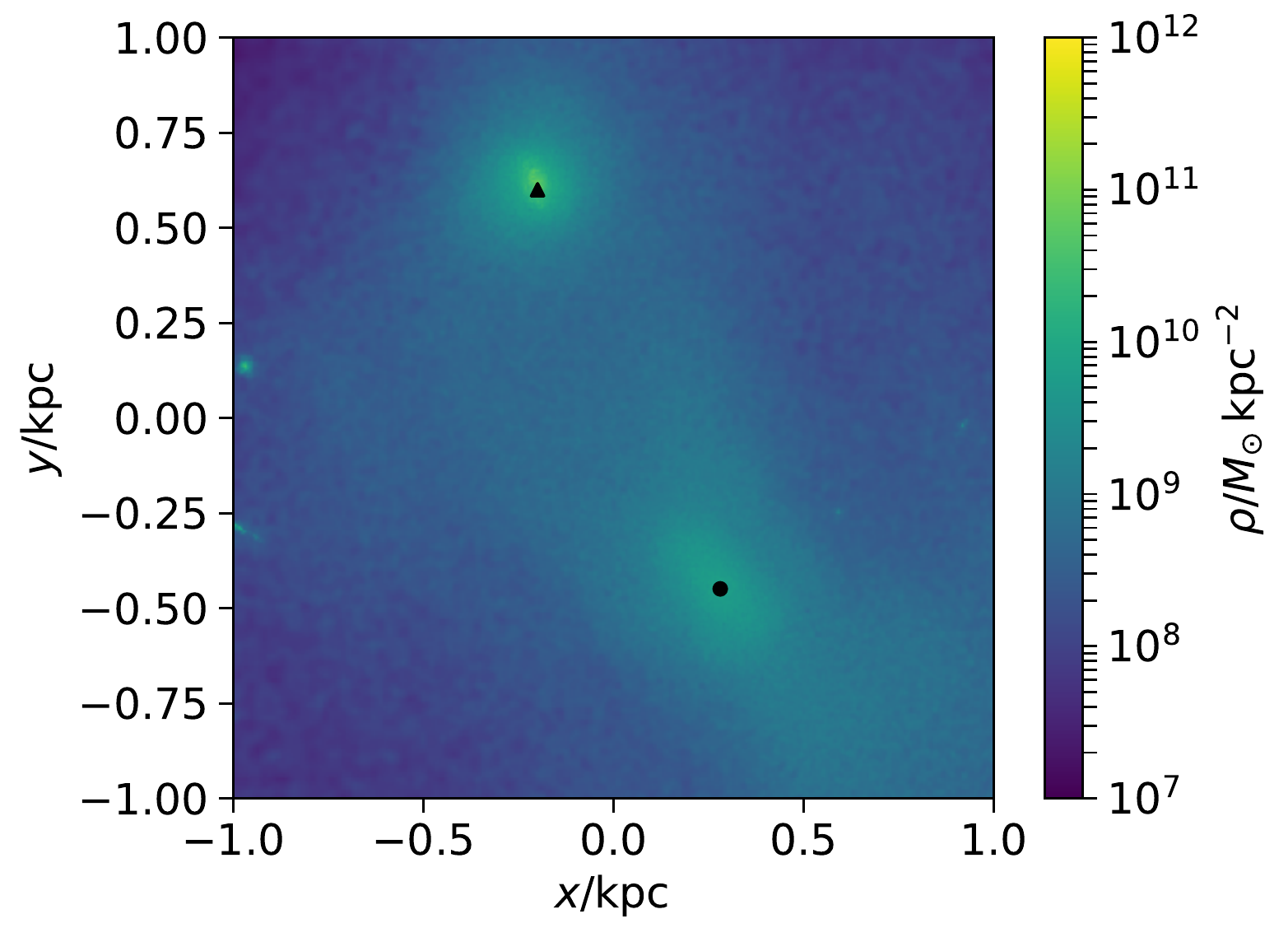}\hfill 
\includegraphics[height=0.22\textheight]{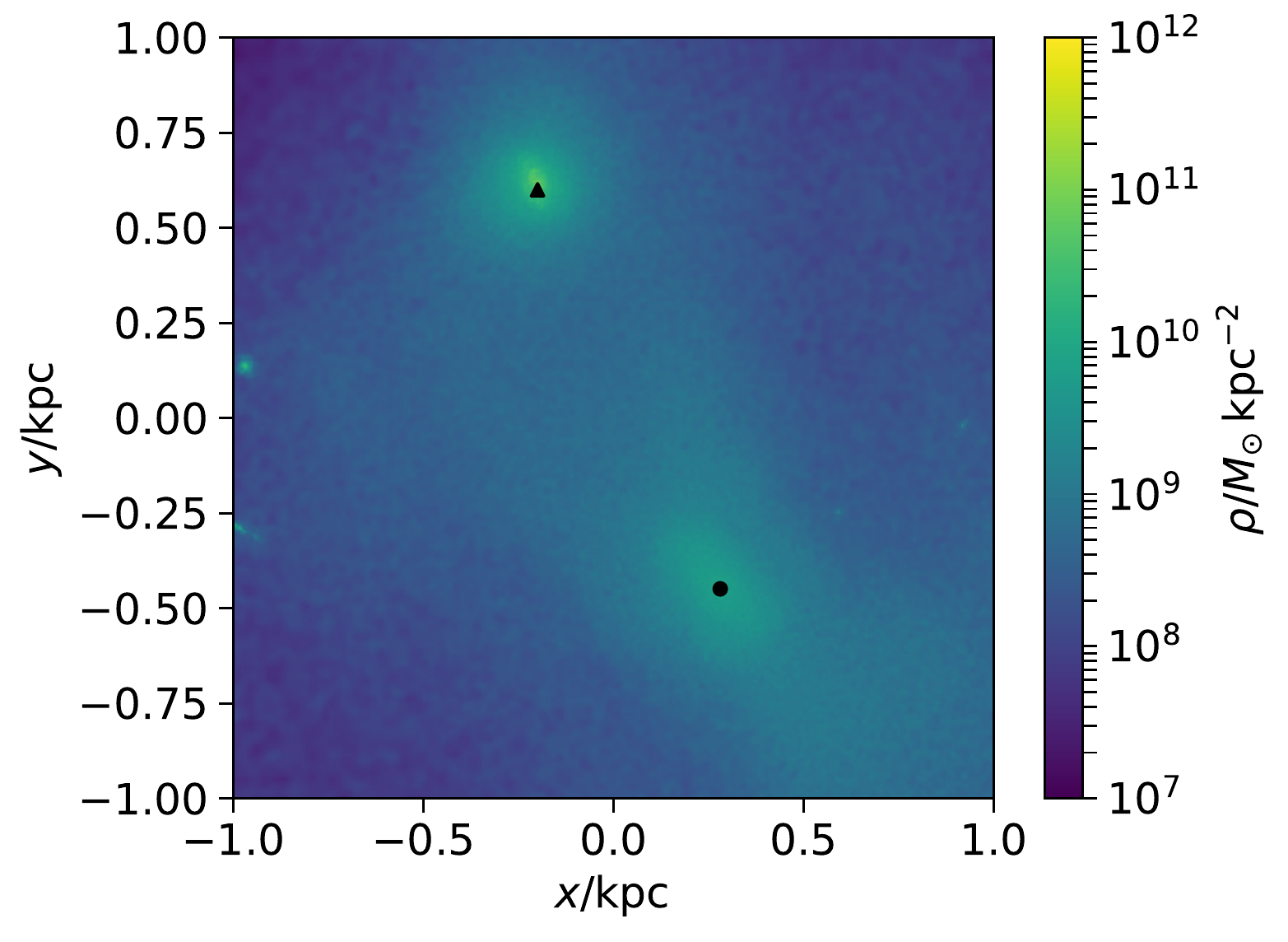}
\includegraphics[height=0.22\textheight]{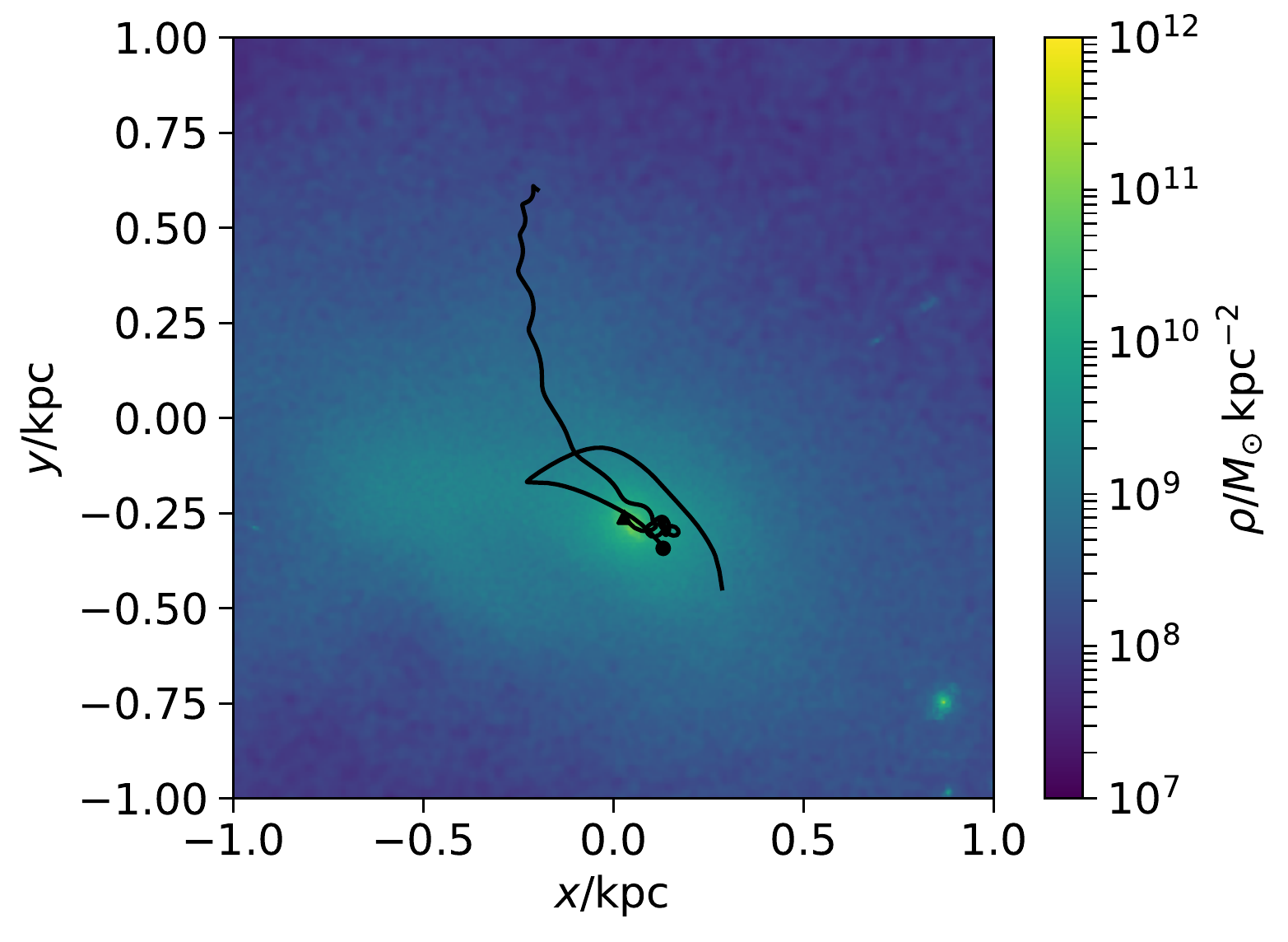}\hfill 
\includegraphics[height=0.22\textheight]{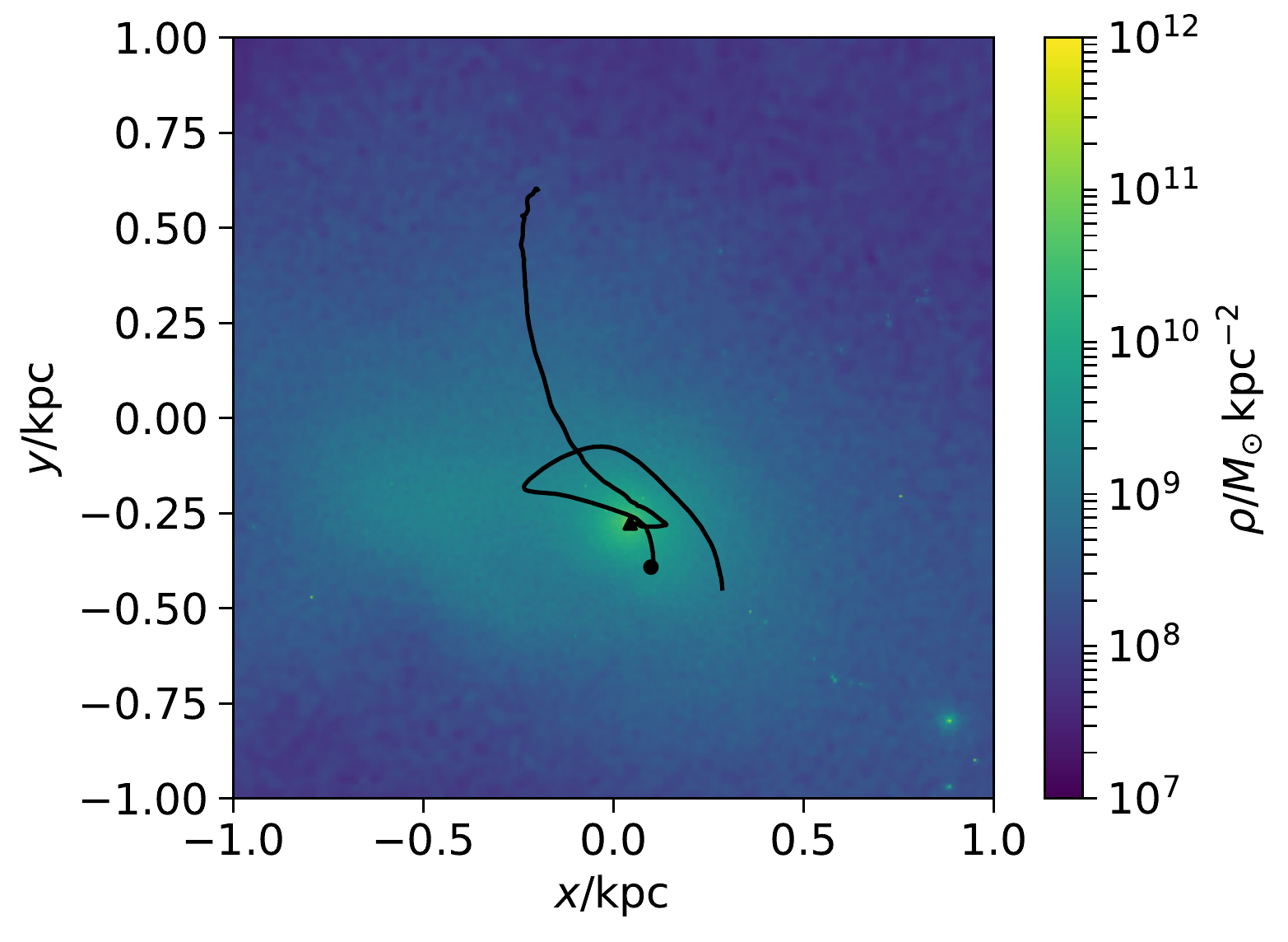}
\includegraphics[height=0.22\textheight]{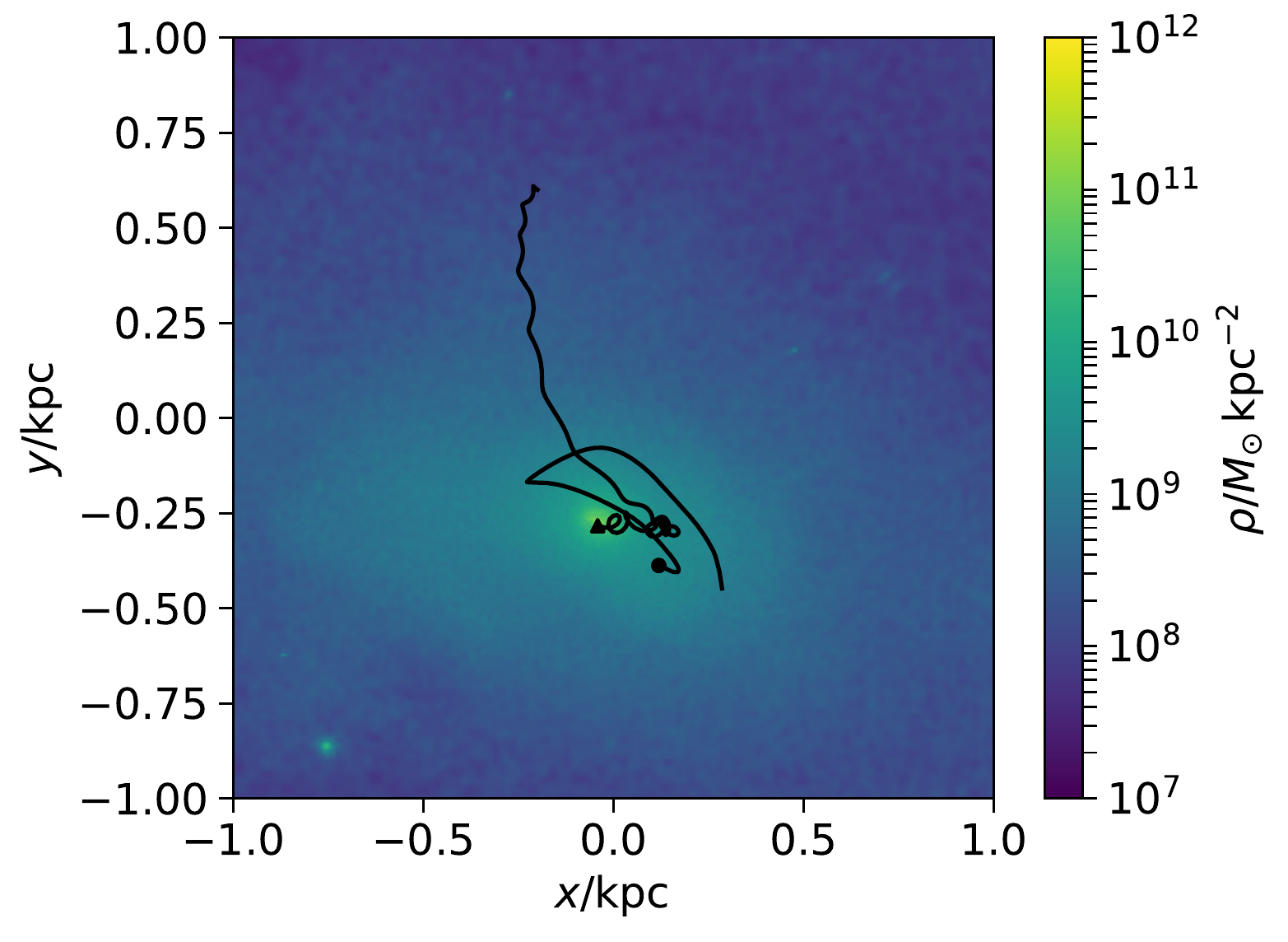}\hfill 
\includegraphics[height=0.22\textheight]{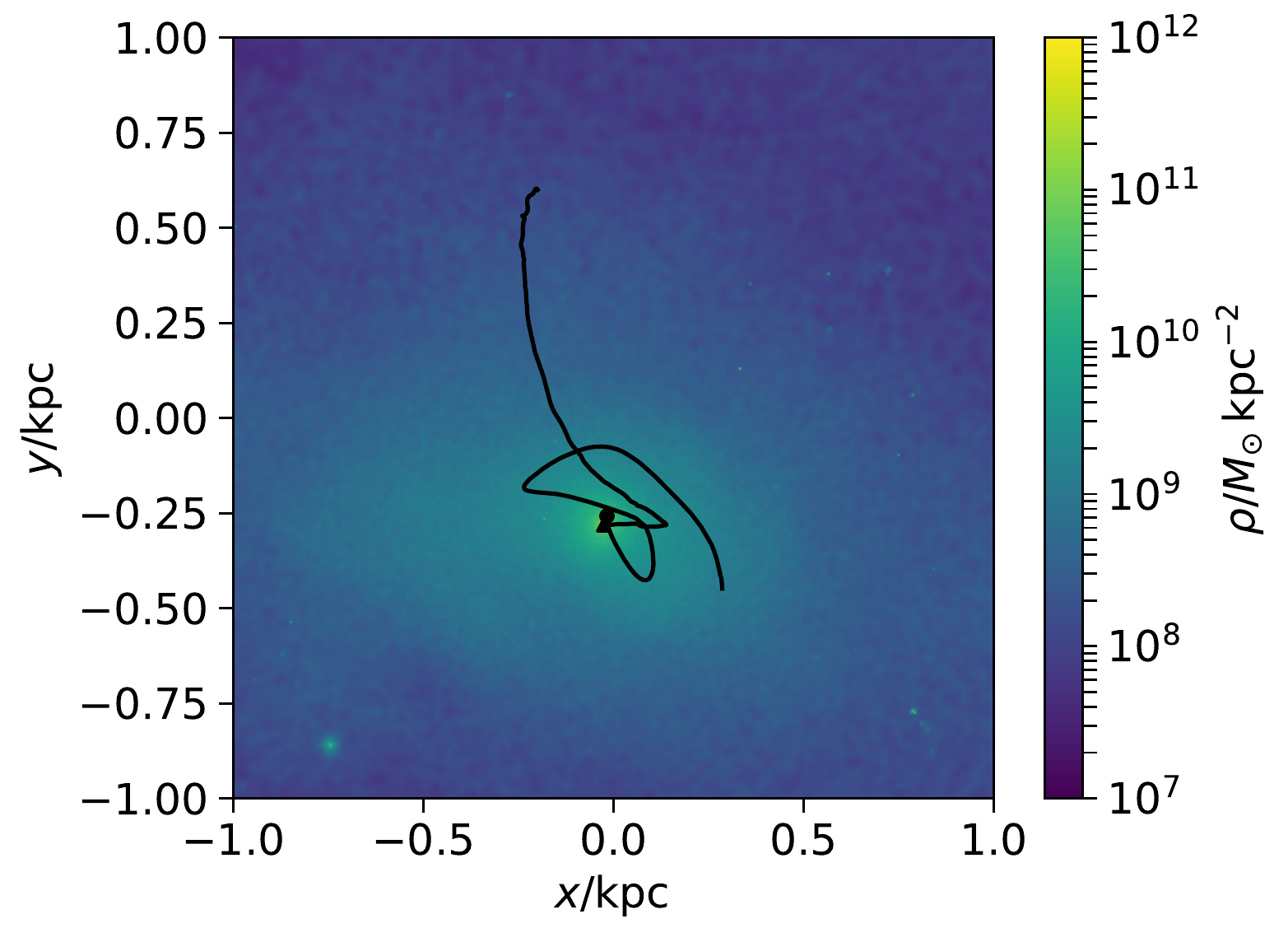}
\includegraphics[height=0.22\textheight]{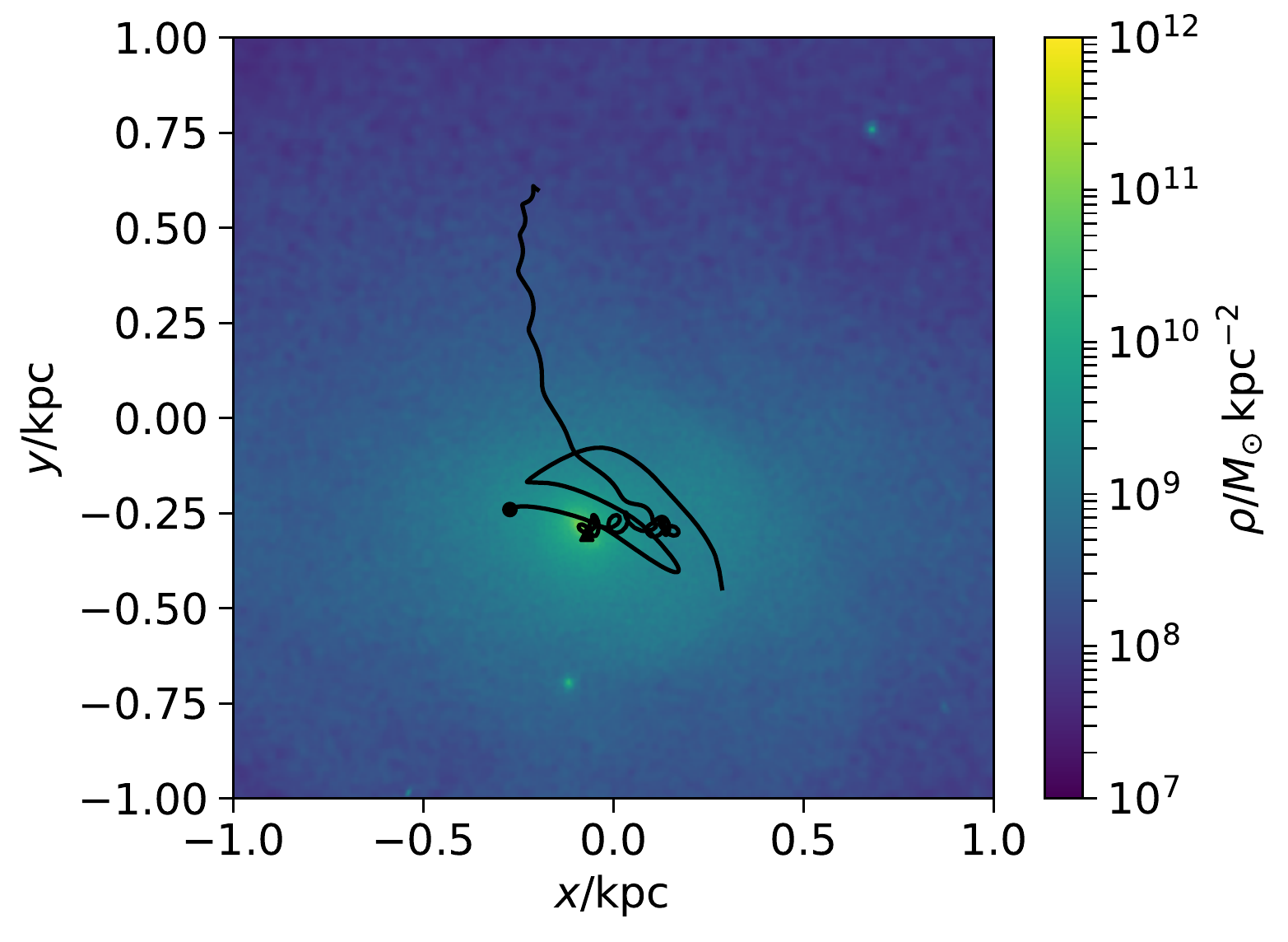}\hfill 
\includegraphics[height=0.22\textheight]{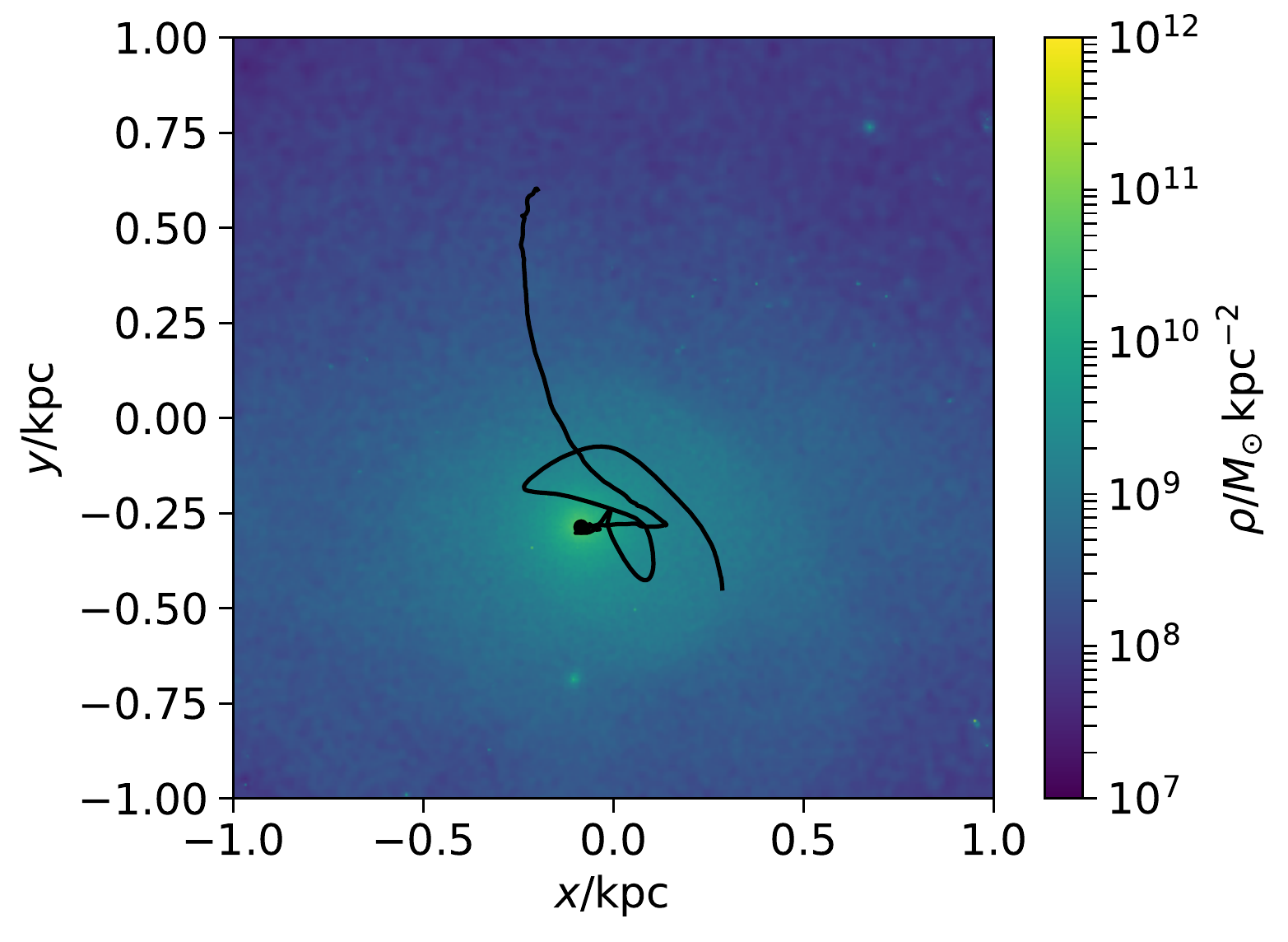}
\caption{Stellar surface density maps at, from top to bottom, $t=0, 15, 17$, and $20\Myr$. The left panels represent R20, whereas the right panels represent R1. Initially, BH2 (triangle), at ($x$,$y)\sim(-0.3,0.6$) kpc, is surrounded by a dense stellar nucleus (N2), while the more massive BH1 is surrounded by a shallow stellar distribution (what remains of N1 after tidal shocks have affected it). At $t=15\Myr$, BH1 approaches N2 and finds itself in a much denser stellar distribution that speeds up dynamical friction.}
 \label{fig:StarMaps}
\end{figure*}

In this Section, we study how well-resolved density and gravity allow us to understand the dynamical evolution of SMBHs.

We show here that, in principle, dynamical friction is sufficient to explain the decay of the orbiting SMBH. We calculate how the specific angular momentum, $L=rv$ in the case of circular orbits, where $r$ is the distance between the two SMBHs and $v$ the relative velocity, varies with time. We assume that the object moves on circular orbits and feels dynamical friction from a uniform background with a density varying with time. This means that the specific angular momentum varies according to the following equation:
\begin{eqnarray}
\frac{\d L}{\d t}=-4\pi \ln(\Lambda)\G^2M_\text{BH1}\frac{\rho(t)r(t)}{v(t)^2}\left[ \text{erf}(X)-\frac{2X}{\sqrt{\pi}}e^{-X^2}\right],\ \label{eq:DynamicalFriction}
\end{eqnarray}
where $X=v/\sqrt{2}\sigma$ and $\sigma$ is the velocity dispersion around the orbiting object, set equal to $150 \km \s^{-1}$ by fitting the density profile of N2 with an isothermal sphere. $M_\text{BH1}$, which varies by less than 10 per cent during the simulation, is set to its value at $t_0$, i.e. $6.5\times10^6\Mdot$.

The quantities $v$ and $r$ are direct outputs from the simulation, whereas $\rho(t)$ is estimated from the spherical total density profile centred on BH2, at the instantaneous location, $r$, of BH1 at each timestep. The evolution of the density ``seen" by BH1 is shown in the top panel of Fig.~\ref{Fig:RhoAndV}, for all resolutions. We find that all densities agree very well for $t<13$~Myr, which is consistent with the orbits of BH1 being the same, regardless of resolution. The runs give different results at later times, as expected, with R1 and R2 more similar out to $t=17$~Myr, and R5 and R20 differing substantially. We will return to the density evolution below.

We chose the value of the Coulomb Logarithm \citep{BT_87} as $\ln(\Lambda) = \ln\left(b_{\text{max}}\sigma^2/\G(M_\text{BH1}+m)\right) \sim 10$, where $m\sim5\times10^3\Mdot$ is the average mass of particles in the simulations and the chosen value of 2~kpc for $b_{\text{max}}$ is rather arbitrary but does not strongly impact the final result. Finally, for the initial conditions, we take the value of the specific angular momentum from the simulation at $t_0$.

In the bottom panel of Fig.~\ref{Fig:RhoAndV}, we show the value of the specific angular momentum as a function of time obtained through the numerical integration of Eq.~(\ref{eq:DynamicalFriction}), together with the actual value in the simulation. We insist on the fact that $L$ is \emph{not} implemented as such in the simulation, where the total force is computed using a multipole hierarchical method.

In the high-resolution cases, both in the model and in the simulation, we observe a sharp decay at $\sim 17\Myr$, meaning that the loss of angular momentum is sufficient to make the two bodies get close and form a binary. At this point, we stop the integration of Eq.~(\ref{eq:DynamicalFriction}) because our model is not valid anymore: when the binary is formed, the dynamics is mainly driven by single interactions between the two SMBHs and not by dynamical friction. Of course there are differences between our model and the simulations but this is expected since the dynamics is not only driven by dynamical friction but also by local variations of the potential, and overall the matching is acceptable.

In Fig.~\ref{fig:StarMaps}, we show maps of the stellar surface density at different times, and the different trajectories of the two SMBHs for R20 and R1. Comparing R1 and R20, we see that in R20, where the gas softening is 20 pc, BH1 does not ``stick'' to N2, i.e. the gravitational interaction is not sufficiently well resolved for BH1 to be captured in the dense stellar nucleus where dynamical friction can be effective. The same occurs in R5 (cf. Fig. \ref{fig:DistBHvsTime}). The stellar, gas, and dark matter density profiles in all the simulations at $t = 17$ Myr, i.e. at the moment when BH1 merges into N2 in the R1 and R2 runs (but not in R5 and R20), are shown in Fig. \ref{fig:DensityAllComponents}. The local density around the orbiting BH is the same for all the runs, showing that in the passage at 17 Myr the effect of dynamical friction is not enhanced by a higher stellar density in R1 and R2. This is also verified in Fig. \ref{Fig:RhoAndV}, where the local density at the position of the orbiting BH is shown as a function of time. Notably, at the pericentric passage at $t=13$ Myr, the density in R5 is slightly higher than in R1 and R2. Notwithstanding, BH1 in R5 is not dragged faster towards BH2.  BH1 finds itself in a high density region at $t = 13$ Myr, with densities similar to R1 and R2, but it then moves out of the nucleus of BH2 and the surrounding density decreases. R2 behaves similarly to R1, while the behaviour of BH1 in R5 is similar to R20’s: the SMBH passes through a high-density region, but the gravitational force is not sufficiently well resolved. In the case of R1 and R2, where the force is better resolved, BH1 is quickly caught by N2, whereas in the case of R5 it passes through the nucleus and gets caught at a later time.



This gives us the following criterion: to be able, in numerical simulations, to capture the formation of the SMBHB, dynamical friction must be well resolved, meaning that the wake lagging the orbiting BH must be resolved, and that spatial resolution must capture the local variation of density, up to scales comparable to the influence radius, $r_\text{inf}$, of BH1:

\begin{eqnarray}
r_\text{inf}=\G M_\text{BH1}/\sigma^2 \, .
\label{eq.RadiusInfluence}
\end{eqnarray}
For $M_\text{BH1}=6.5\times10^6\Mdot$ and $\sigma=150\km \s^{-1}$, we have $r_\text{inf} \simeq 1\pc$, which explains why we capture the dynamics well at pc resolution but not with 5 or 20 pc.

To conclude, we showed here that dynamical friction \emph{is} efficient enough to explain the sharp decay of angular momentum and, consequently, of the distance between the two SMBHs, from kpc to pc scales, down to the formation of a SMBHB. However, to properly capture the dynamics in numerical simulations, the radius of influence of BHs must be resolved. We also found that, for this merger, not all the components have the same role: the density in the smooth gas component around BH1 is much lower than the stellar one, hence its contribution to dynamical friction is negligible. The simulated galaxies were fairly gas-rich at the beginning of the simulation (30 per cent of gas in the disc), but much of this gas was consumed in SF during the early phases of the merger. High-redshift galaxies can have a much higher gas fraction, $\sim 50-60$ per cent \citep{Tacconi_10}, but a central starburst would decrease the gas fraction as well. For reference, in the merger with 60 per cent gas fraction in the suite of \cite{Capelo_15}, the final gas fraction within 1 kpc was $\simeq 20$ per cent after the starburst. In galaxies with an even higher gas fraction, or with more inefficient SF, gas is likely to play a more important role.

\begin{figure}
 \includegraphics[width=\columnwidth]{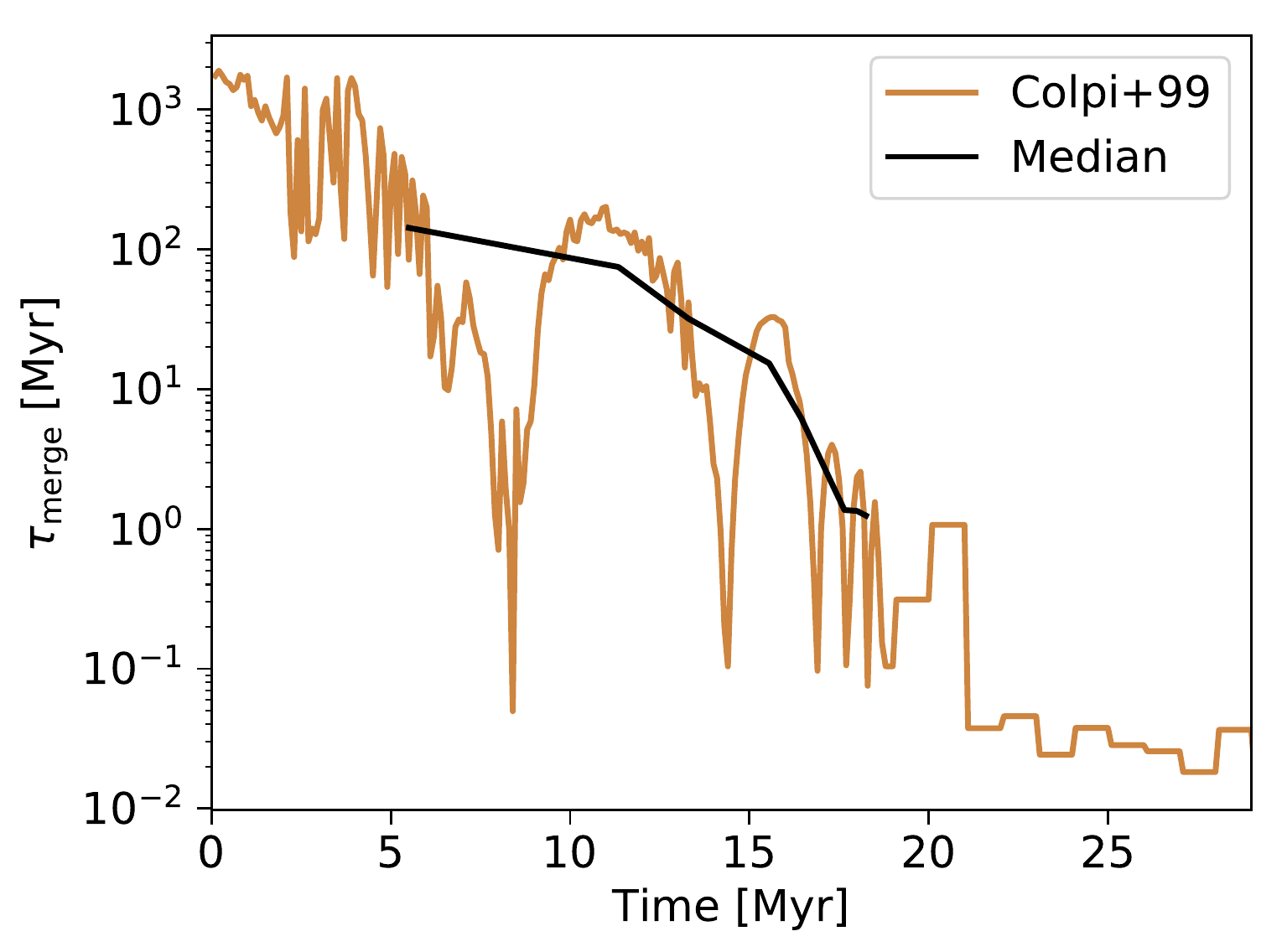}
 \caption{For simulation R1, we show $\tau_\text{merge}$ at each timestep, according to Eq.~(\ref{eq:Lacey}) and its value when computed using the median value of each quantity within one orbit.}
 \label{fig:TmergevsTime}
\end{figure}

\subsection{Analytical models and merger time-scales}
In simulations with lower resolution, or in semi-analytical models, analytical expressions for merger time-scales are often used to estimate the time needed for SMBHs to form a binary or coalesce. To provide a benchmark, we compare our numerical result to an analytical estimate of the time needed for a satellite to merge within a larger system. \cite{Lacey_93}  and \cite{Colpi_99} estimate analytically the time needed for a satellite halo to merge with a more massive halo. While they derive their equation for a satellite orbiting inside a fixed halo, we assume that it is still valid when the satellite (here the SMBH) moves in a stellar bulge that is itself moving. Defining $\tau_\text{merge}$ as the time to merge after $t_0$, the generic form of their equation can be written as:
\begin{eqnarray}
\tau_\text{merge}=1.17\frac{r^2_\text{circ}V_\text{circ}}{\G M_\text{sat}\ln(M_\text{enc}/M_\text{sat})}\varepsilon^{\alpha} \text{ ,}\label{eq:Lacey}
\end{eqnarray}
where $r_\text{circ}$ is the radius of the circular orbit having the same energy of the actual orbit (hereafer circular orbit), $V_\text{circ}$ is the speed of a satellite on the circular orbit, $M_\text{sat}$ is the mass of the satellite, in our case the mass of the orbiting SMBH, $M_\text{enc}$ is the total enclosed mass\footnote{In the original papers, the authors suggest to take the mass of the halo, $M_\text{halo}$, instead of the enclosed mass $M_\text{enc}$. We checked that this does not strongly impact the results since this quantity is taken in a logarithm. We adopt this convention because it does not depend on the definition of a \emph{halo}.} in a sphere centred on the central SMBH of radius $r_\text{circ}$, and the circularity $\varepsilon=J(E)/J_\text{circ}(E)$ is the ratio between the angular momentum and the one corresponding to a circular orbit. \cite{Lacey_93} suggest $\alpha=0.78$, whereas \cite{Colpi_99} suggest $\alpha\in[0.4;0.78]$ depending on if the orbit is cosmologically relevant or not. Further, we assume that $M_\text{sat}=M_\text{BH1}$, which is not correct initially, since, even though there has been a nuclear coup, BH1 is not completely naked and one should take into account how the remnant of the stellar nucleus around BH1 evolves with time (see, for instance, the top panels in Fig. \ref{fig:StarMaps}). Since we use this formalism only as a reference, and our main interest is to study which processes affect the dynamics, rather than giving a precise time-scale, we do not modify Eq. (\ref{eq:Lacey}), but in the simulation the ``real" mass of the satellite is stars+SMBH, at least initially \citep{Yu2002}. Finally, $\tau_\text{merge}$ is defined as the time needed for the merger whereas the simulations we present in this paper have the resolution to capture the formation of the binary, which occurs before the merger. All these effects result in a $\tau_\text{merge}$ which will typically be larger than $\tau_\text{SMBHB}$.

We calculate $\tau_\text{merge}$  at each timestep in simulation R1 from Eq.~(\ref{eq:Lacey}) and show it in Fig.~\ref{fig:TmergevsTime}, for $\alpha=0.4$. As one can see, instantaneously, the time-scales calculated with this formalism have large variations: at an apocentre or the following pericentre, the time-scale can vary by more than four orders of magnitude, with the time-scale longer at apocentre and shorter at pericentre. If one were to ``add" this time-scale at the end-point of a low-resolution simulation, using just the information at one timestep can lead to widely different results. A better approach is to use a more stable value. 

We show $\widetilde{\tau}_\text{merge}$ using $\widetilde{r}_\text{circ}$, $\widetilde{V}_\text{circ}$, $\widetilde{M}_\text{enc}$, and $\widetilde{\epsilon}$, where the tilde indicates the median value within one orbit, from one pericentre (apocentre) to the following of each quantity. Using this quantity, we recover more reasonable time-scales, although still larger than the time needed for the formation of the binary, starting from the same position and time, as expected for the reasons described above. We can obtain a lower limit rescaling $\widetilde{\tau}_\text{merge}$ by the ratio between the stellar mass enclosed within 100~pc from BH1 and the SMBH mass at the beginning of the simulation. This ratio is $ \sim 60$,  and the binary should form after 2.4 Myr if the nucleus remained intact throughout the evolution. In conclusion, if one wants to estimate the merger time-scale from dynamical friction, at the end of a low-resolution simulation for instance,  taking the median value of the last orbit rather than simply computing the value for the last output  gives more accurate and stable results, with the evolution of the remnant of the stellar nucleus around the satellite BH bracketing lower and upper limits. 

\begin{figure}
	\includegraphics[width=\columnwidth]{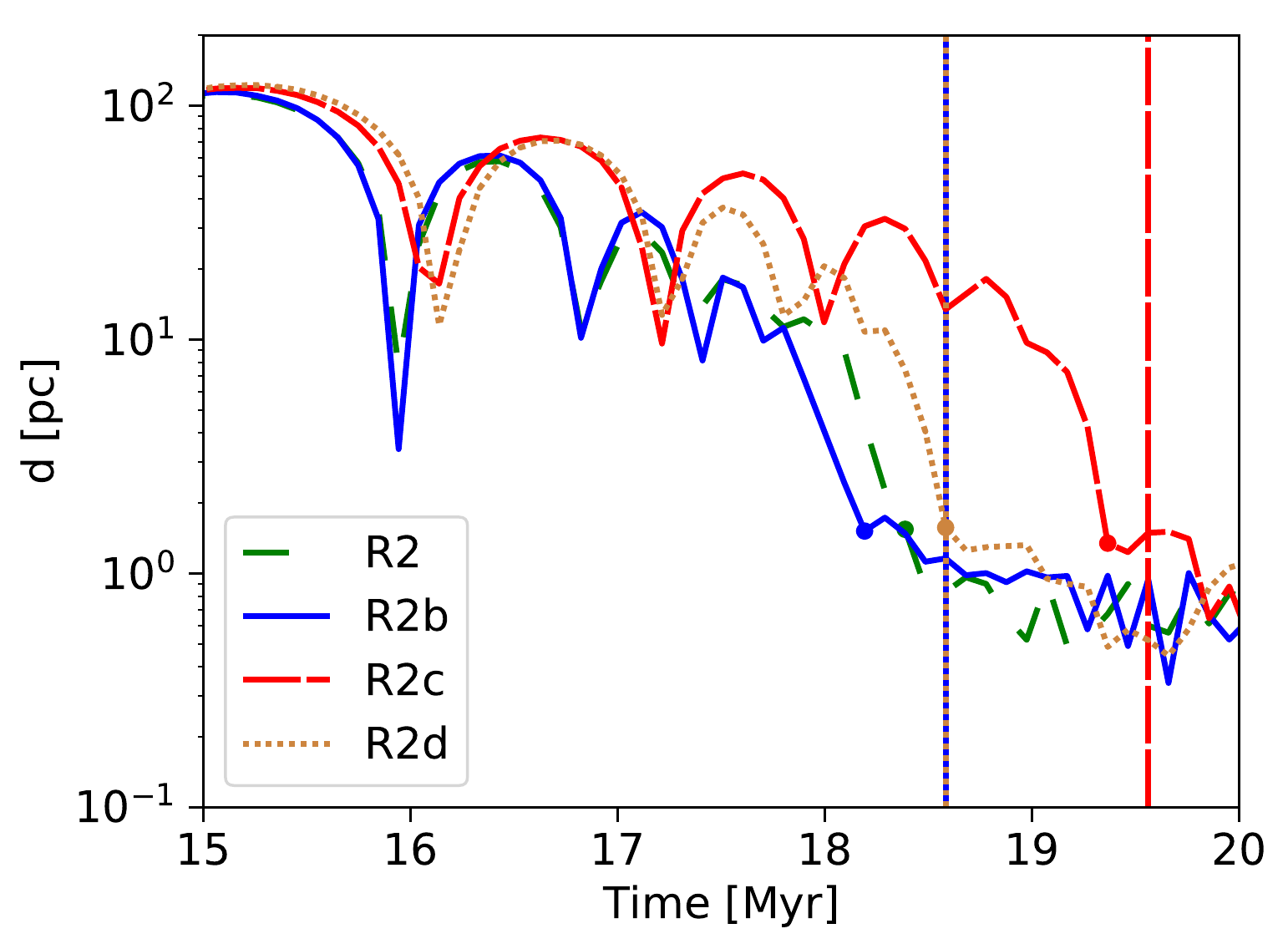}
	\caption{Distance between the two SMBHs as a function of time for different runs of same resolution but with slightly different initial conditions. The vertical lines show when the SMBHB is formed. The dots indicate the first time the distance between the two SMBHs is below resolution (2~pc).}
	\label{fig:BHDistShift}
\end{figure}
\section{Stochasticity of the trajectory}
\label{StochasticityOfTheTrajectory}
In this Section, we study how  our results depend on the initial conditions. As the resolution increases and denser gas and stellar clumps can be resolved, one may expect that random scatterings with a perturber may affect the SMBH's trajectory. In Section~\ref{ShiftingBlackHoles}, we vary the initial parameters of one of our simulations (R2) to see if the dynamics is affected. In Section~\ref{EffectsOfGasClumps}, we quantify the effects of gas clumps.


\subsection{Shifting black holes}
\label{ShiftingBlackHoles}

We have seen in Section~\ref{DynamicalEvolutionOfBHs} that the dynamics of  SMBHs can be understood looking at the smooth stellar potential in the nucleus. However, the gas map at the same scales shows gaseous clumps that could, in principle, scatter SMBHs, inducing random motions. To assess the relevance of perturbations caused by clumps, we slightly change the orbital parameters of the SMBHs 12 Myr after our R2 simulation has begun. Either the position of BH1 is shifted by 3 (Run2b) or 16 (Run2d) pc, at fixed separation from BH2, or the speed of BH2 is increased by 20 per cent (Run2c).

We show our results in Fig.~\ref{fig:BHDistShift}. All cases are very similar: we observe the same sharp decrease of the distance between SMBHs at $\sim18\Myr$; this decrease occurs when the SMBHB forms and just before the gravitational resolution is reached. We conclude therefore that the SMBH trajectory is not significantly affected by discrete perturbers. We show in Section~\ref{EffectsOfGasClumps} why clumps of material, which are resolved in higher-resolution runs and not in the initial simulation, do not play a relevant role in the dynamics of  SMBHs in our galaxies.

\subsection{Effects of gas clumps}
\label{EffectsOfGasClumps}
\begin{figure}
	\includegraphics[width=\columnwidth]{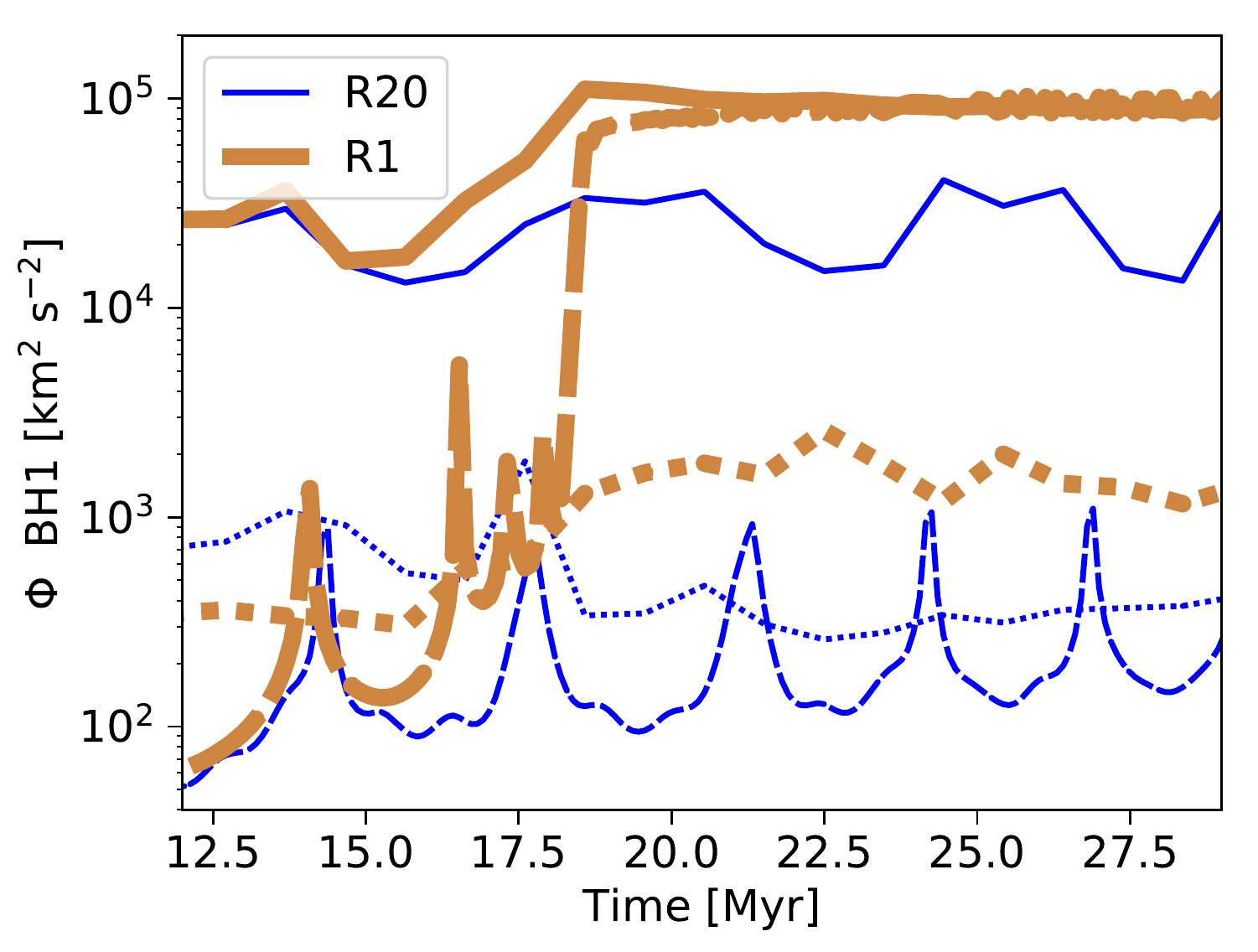}
	\caption{Potential felt by BH1 due to stars within 1~kpc from BH1 (solid line), due to BH2 (dashed lines), and due to gas clumps within 1~kpc (dotted line). All quantities are plotted as a function of time.}
	\label{fig:BH1Potential}
\end{figure}

In this Section, we study the effects of gas clumps on the dynamical evolution of SMBHs. We use the clump finder {\scshape skid}\footnote{Freely available at \url{https://github.com/N-BodyShop/skid}.} \citep{Stadel_01} to identify all the gas clumps within 1 kpc from BH1. We focus on gas because we have found the gas density, in contrast to the smooth stellar density, to be clumpy. The gas clumps have masses between a few times $10^4\Mdot$ and $10^6 \Mdot$. The clumps' mean gas density distribution peaks at  $\sim 10^2$ particles cm$^{-3}$, which is in very good agreement with the typical densities of giant molecular clouds  \citep{1999ASIC..540...29M}, with only a small tail at higher densities (the mass fraction in gas with density $>10^4$ particles cm$^{-3}$ is $<10\%$). The gas density is always below the ``effective density of the SMBH", defined as that the SMBH mass would have if spread over a sphere with radius the softening length, therefore we are not affected by spurious motions \citep{2015ApJ...811...59D,SouzaLima_17}.  Our simulation is also unaffected by an over-estimate of stochastic gravitational interactions with over-dense gas clumps \citep{2015ApJ...811...59D,SouzaLima_17}. 

In Fig.~\ref{fig:BH1Potential}, we compare the potential felt by BH1 due to stars within 1 kpc from BH1, due to gas clumps in the same region, and due to BH2. We show that, at the moment the SMBHs form a SMBHB, BH2 becomes an important source of potential, as important as all the stars within 1 kpc around the binary. This reflects the criterion we used to identify a bound binary, where the potential of one SMBH on the other becomes dominant. Even if we observe a higher potential due to clumps when increasing resolution, meaning that more clumps have  formed, the potential remains at least one order of magnitude below the stellar potential and can therefore be neglected.

In Section~\ref{ShiftingBlackHoles}, we changed the orbital configuration of SMBHs to see if the dynamics was driven by interaction with gas clumps. The negative result we obtained, coupled with the analysis of the gaseous potential, confirms that there is no dominant SMBH-SMBH-Clump+background interaction but only SMBH-SMBH+background, and in particular SMBH-SMBH+stellar background interaction.


\begin{figure}
	\includegraphics[width=\columnwidth]{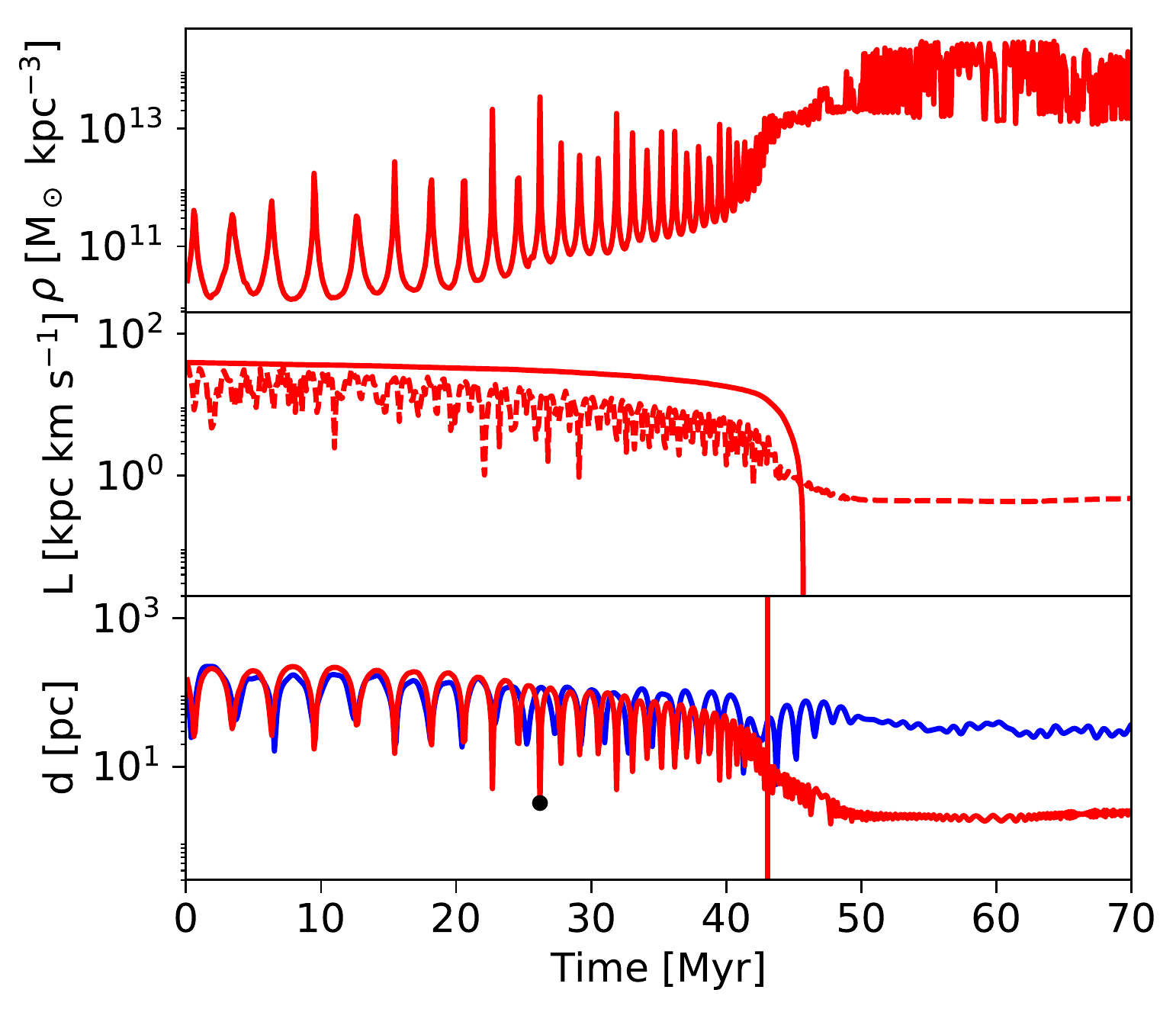}
	\caption{Quantities, as a function of time, for the zoomed-in 1:2 coplanar, prograde--prograde merger (red). We also show the SMBH separation from the 20 pc simulation from \protect\cite{Capelo_15} (blue). Top panel: density at the position of the orbiting SMBH, BH2. Middle panel: specific angular momentum obtained with Eq.~(\ref{eq:DynamicalFriction}) (solid) and from the simulation (dotted). Bottom panel: distance between the two SMBHs. The vertical line indicates the time the SMBHB is formed following our criteria given in Section~\ref{Initial_simulation} and the dot indicates the first time we reach minimal resolution, 5 pc in this case.}
	\label{fig:1to2datas}
\end{figure}

\section{Other mergers}
\label{OtherMergers}
In this Section, we describe our results for two other zoomed-in simulations from \cite{Capelo_15}. With respect to the original simulation, the first one differs only by the mass ratio (1:2 instead of 1:4), whereas the second one differs only by the inclination of the primary galaxy (inclined instead of coplanar). Moreover, in both simulations, no nuclear coup occurs. This allows us to study the effects of the initial mass ratio of galaxies, of the inclination of the orbit, and of the presence/absence of a nuclear coup. We adopt the same technique to perform these zoom-ins, trimming the outer 20~kpc of the remnant galaxy. For these two runs, we decrease the gravitational softening of the gas to 5~pc. While a resolution of 5 pc is not enough to fully capture the formation of the binary, as detailed in section \ref{AFasterDecay}, we are here mainly interested in testing that dynamical friction from the stellar component can drive the SMBHs from kpc to pc scales when there is not a nuclear coup.


\subsection{1:2 coplanar, prograde--prograde merger}
\label{1to2ProgradeProgradeMerger}

In the 1:2 coplanar, prograde--prograde merger, there is no nuclear coup and the least massive SMBH, with a mass of $\sim3\times10^6\Mdot$, is the orbiting object. We present our results in Fig.~\ref{fig:1to2datas}. The top panel shows the mean density causing dynamical friction on the orbiting SMBH as a function of time. At the end of the simulation, the density is similar to that of the initial simulation with the same resolution (R5):$\sim10^{13-14} \Mdot \kpc^{-3}$. Moreover, since in that simulation the orbiting SMBH is three times lighter than in the original one, we expect a time-scale roughly three times longer. This is close to what we observe, since $\tau_\text{SMBHB}$ is in this case $43\Myr$, whereas in the 1:4 run it is $27\Myr$.

In the middle and bottom panels, we show the loss of angular momentum due to dynamical friction and the distance between the two SMBHs, confirming that the main process that drives  SMBHs to pc scales is dynamical friction from the smooth stellar potential in the nucleus. The overall behaviour is similar to the case of R5. On the one hand, we confirm that the orbital evolution is driven by dynamical friction from the stellar background, as the simulation and analytical models behave in a similar way. On the other hand, the relative force is not well captured because 5 pc is not enough to resolve the wake that causes dynamical friction (see Section~\ref{Explanation:DynamicalFriction}). Additionally, BH2 reaches a separation comparable to the resolution before the formation of the binary, according to our definition.

\subsection{1:4 inclined-primary merger}
\label{1to4InclinedMerger}
\begin{figure}
	\includegraphics[width=\columnwidth]{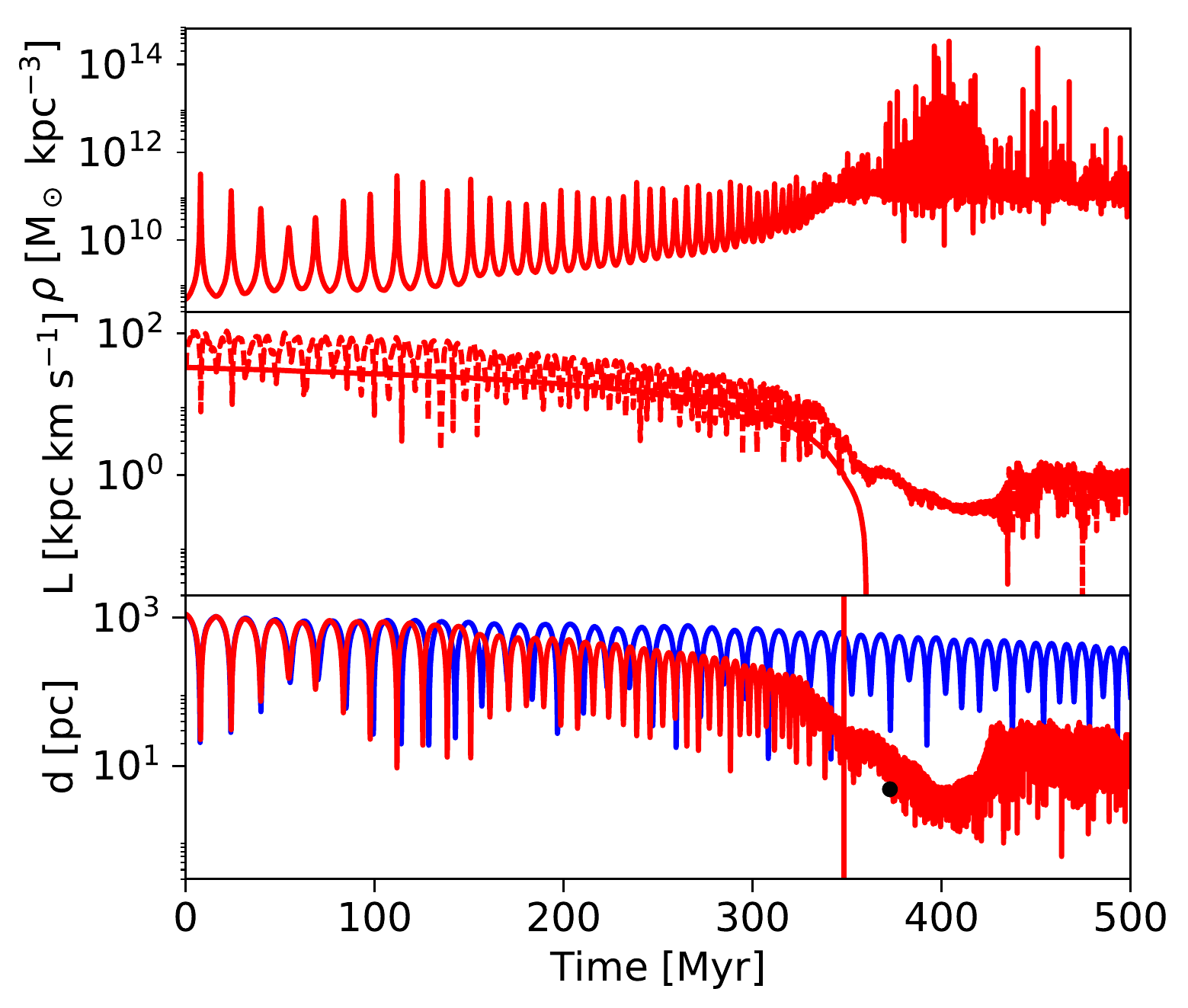}
	\caption{Quantities, as a function of time, for the zoomed-in 1:4 inclined, prograde--prograde merger (red). We also show the SMBH separation from the 20 pc simulation from \protect\cite{Capelo_15} (blue). Top panel: density at the position of the orbiting SMBH, BH2. Middle panel: specific angular momentum obtained with Eq.~(\ref{eq:DynamicalFriction}) (solid) and from the simulation (dotted). Bottom panel: distance between the two SMBHs. The vertical line indicates the time the SMBHB is formed following our criteria given in Section~\ref{Initial_simulation} and the dot indicates the first time we reach minimal resolution, 5 pc in this case.}
	\label{fig:inclineddatas}
\end{figure}
For the 1:4 inclined-primary merger, as the one in the previous Section, there is no nuclear coup: the least massive SMBH, with a mass of $3\times 10^6\Mdot$, is the satellite. Moreover, since the trajectory is inclined instead of coplanar, torques and shocks \citep{Capelo_17} are less efficient at driving gas towards the centre, leading to a lower density in the nucleus. The decay time due to dynamical friction should therefore be longer.
In the top panel of Fig.~\ref{fig:inclineddatas}, we show the mean density causing dynamical friction on the orbiting SMBH, BH2 in this case, which, at the end of the simulation, is lower than in the other cases we have studied: $10^{11}$ instead of $10^{13-14} \Mdot \kpc^{-3}$. A slower decay is confirmed in our simulation, with $\tau_\text{SMBHB}=348$~Myr.

The same considerations discussed for R5 and the 1:2 coplanar, prograde--prograde merger apply. We still observe the loss of angular momentum, confirming that dynamical friction from the smooth stellar component of the nucleus drives the dynamics down to pc scales, but we expect that the ``real" binary formation time-scale is overestimated.

\section{Conclusions}
\label{Conclusions}
We presented a set of zoom-in simulations of already very high-resolution simulations of galaxy mergers. We focus on the dynamics of  SMBHs from the first apocentre, during the merger phase, with a separation smaller than 1.2~kpc, to the formation of the SMBHB. We summarize our findings below:
\begin{itemize}
\item We confirm that the formation of the SMBHB occurs when the two SMBHs are separated by a few pc.
\item We show that dynamical friction from the smooth stellar potential is efficient enough to drive SMBHs from kpc to pc scales. Conversely, neither the gaseous potential nor the dense clumps affect the SMBHs dynamics.
\item We conclude that it is necessary to resolve the influence radius of the orbiting SMBH to be able to capture dynamical friction in the final stages of the merger.
\item  We show that analytical estimates of merger time-scales driven by dynamical friction cannot be computed at a particular moment, especially not at apocentres (pericentres), where the time-scale is overestimated (underestimated). Instead, we suggest to take the median value over one orbit (typically the last one available in the simulation) to have a more accurate result.
\end{itemize}
\section*{Acknowledgments}
We thank Luciano del Valle, Alexander Wagner, and Yohan Dubois for helpful comments and suggestions. We also thank the anonymous referee for giving comments improving the solidity of the numerical approach. This work has been made on the Horizon Cluster, hosted by Institut d’Astrophysique de Paris. We acknowledge Stephane Rouberol for smoothly running the cluster. AL and MV acknowledge support from the European Research Council (Project No. 267117, AL; Project no. 614199, AL, MV). PRC acknowledges support by the Tomalla Foundation.



\bibliographystyle{mnras}
\bibliography{bibli}



\appendix

\section{Effects of trimming the galaxy}
\label{sec:EffectsOfTrimmingTheGalaxy}

\begin{figure}
  \includegraphics[width=\columnwidth]{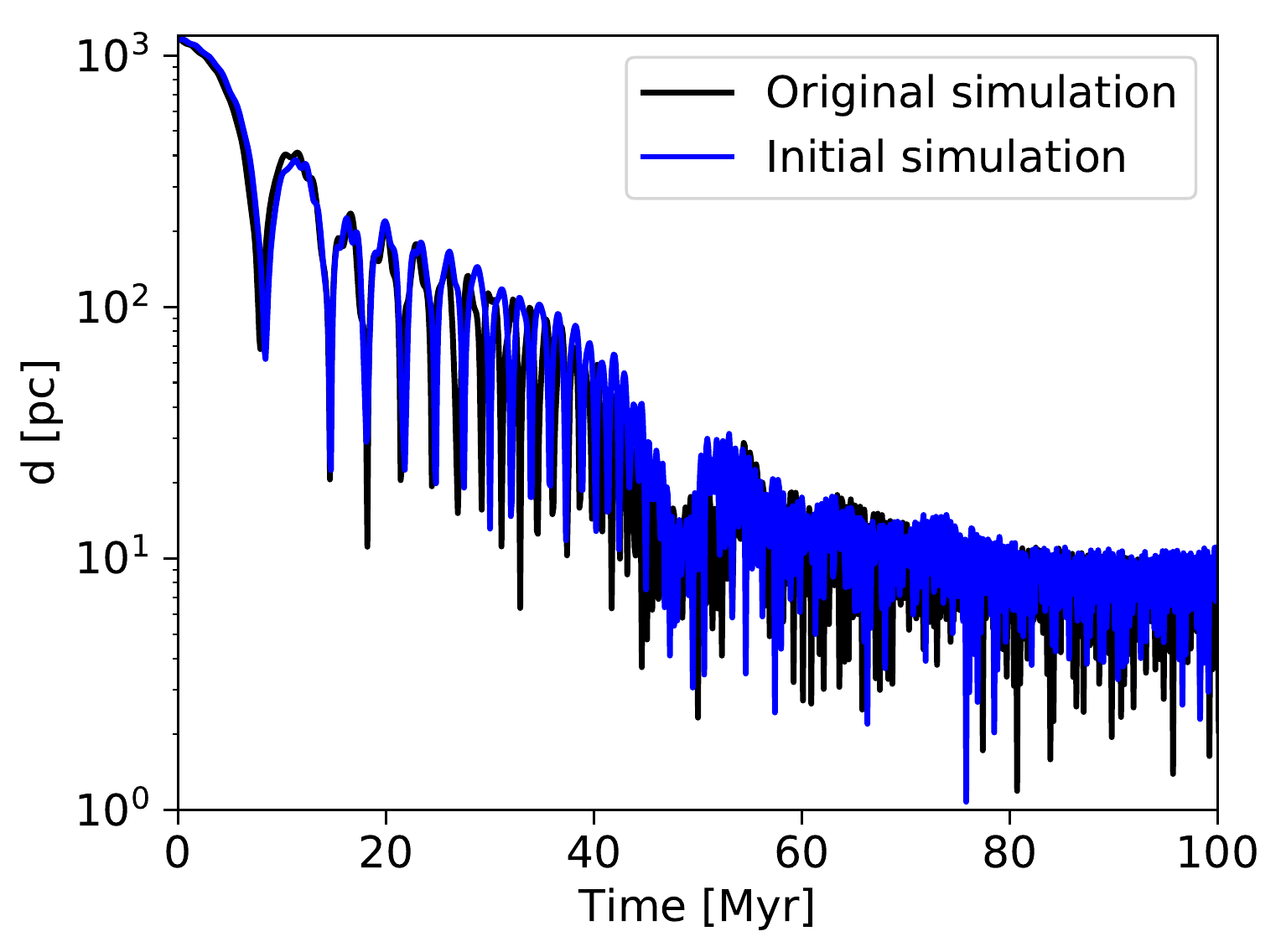}
 \caption{Distance between the two BHs as a function of time, in the simulation from \protect\cite{Capelo_15} (original simulation) and in our simulation where we removed the outer part of the remnant galaxy (initial simulation). We have an excellent agreement during the 100 Myr we simulated, meaning that we can safely trust our results of the increased resolution simulations which are ran for $\sim$30 Myr.}
 \label{DistPedroAndI}
\end{figure}

\begin{figure*}

 \includegraphics[width=\columnwidth]{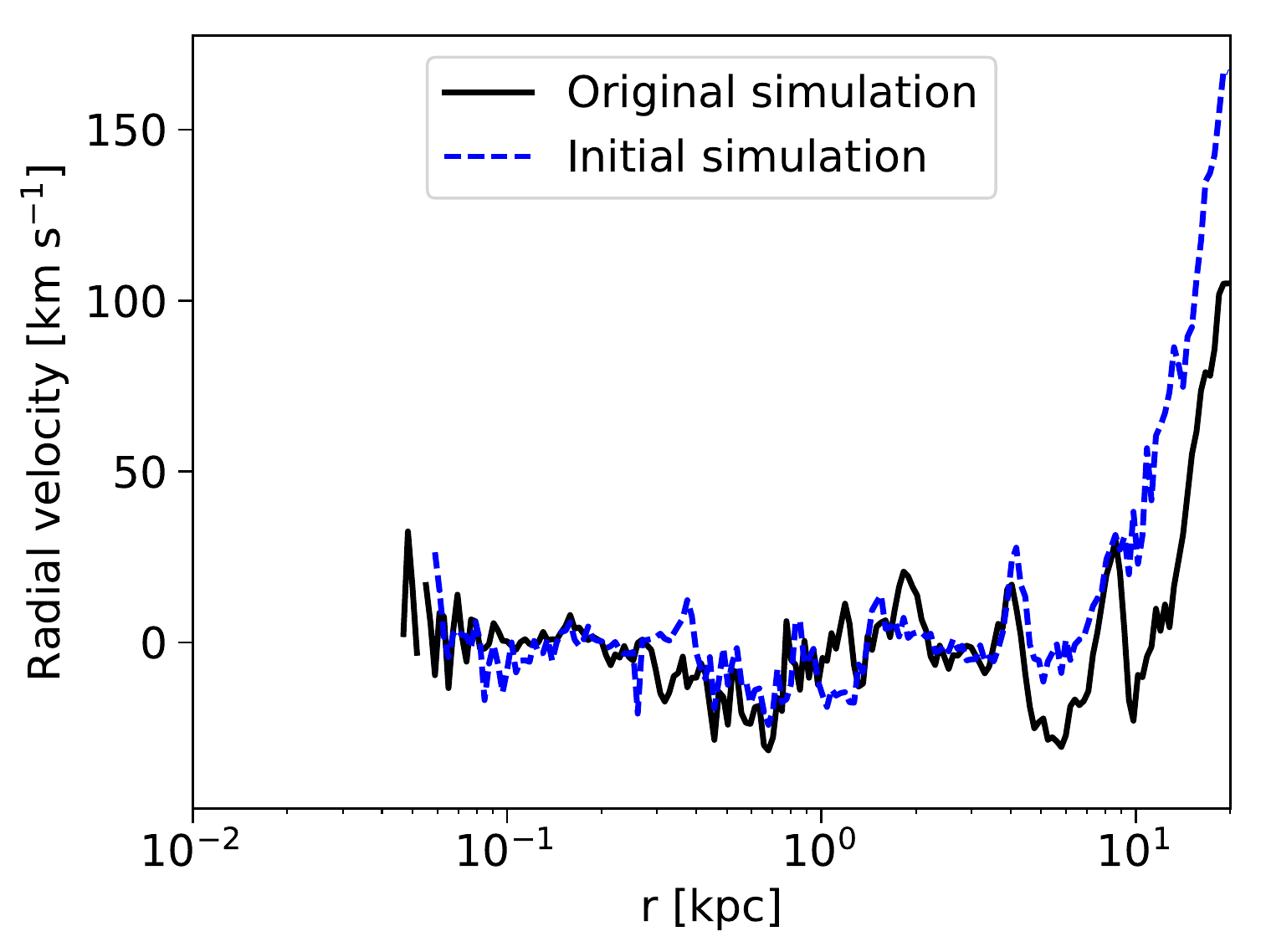}
 \includegraphics[width=\columnwidth]{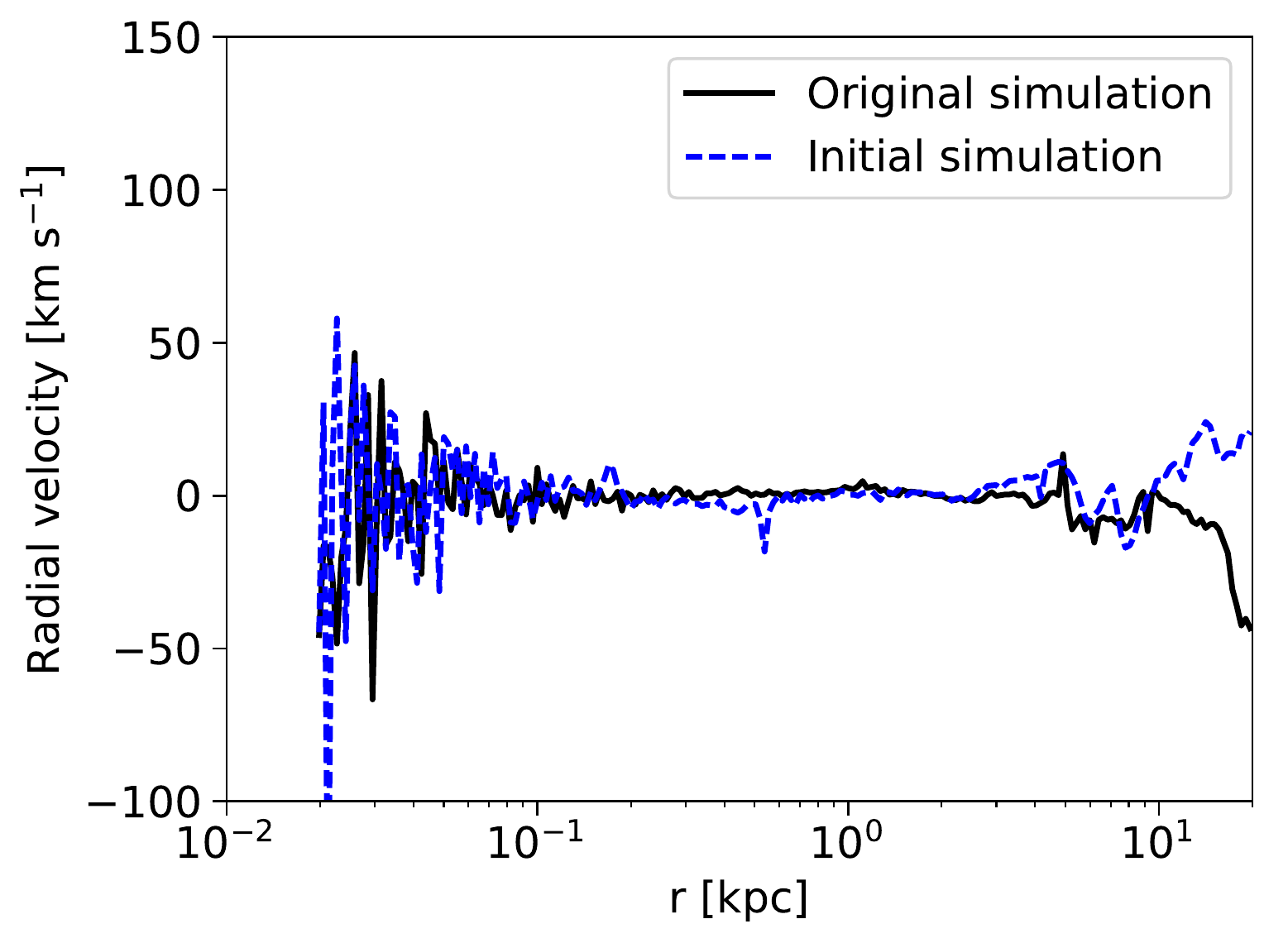}\\
 \includegraphics[width=\columnwidth]{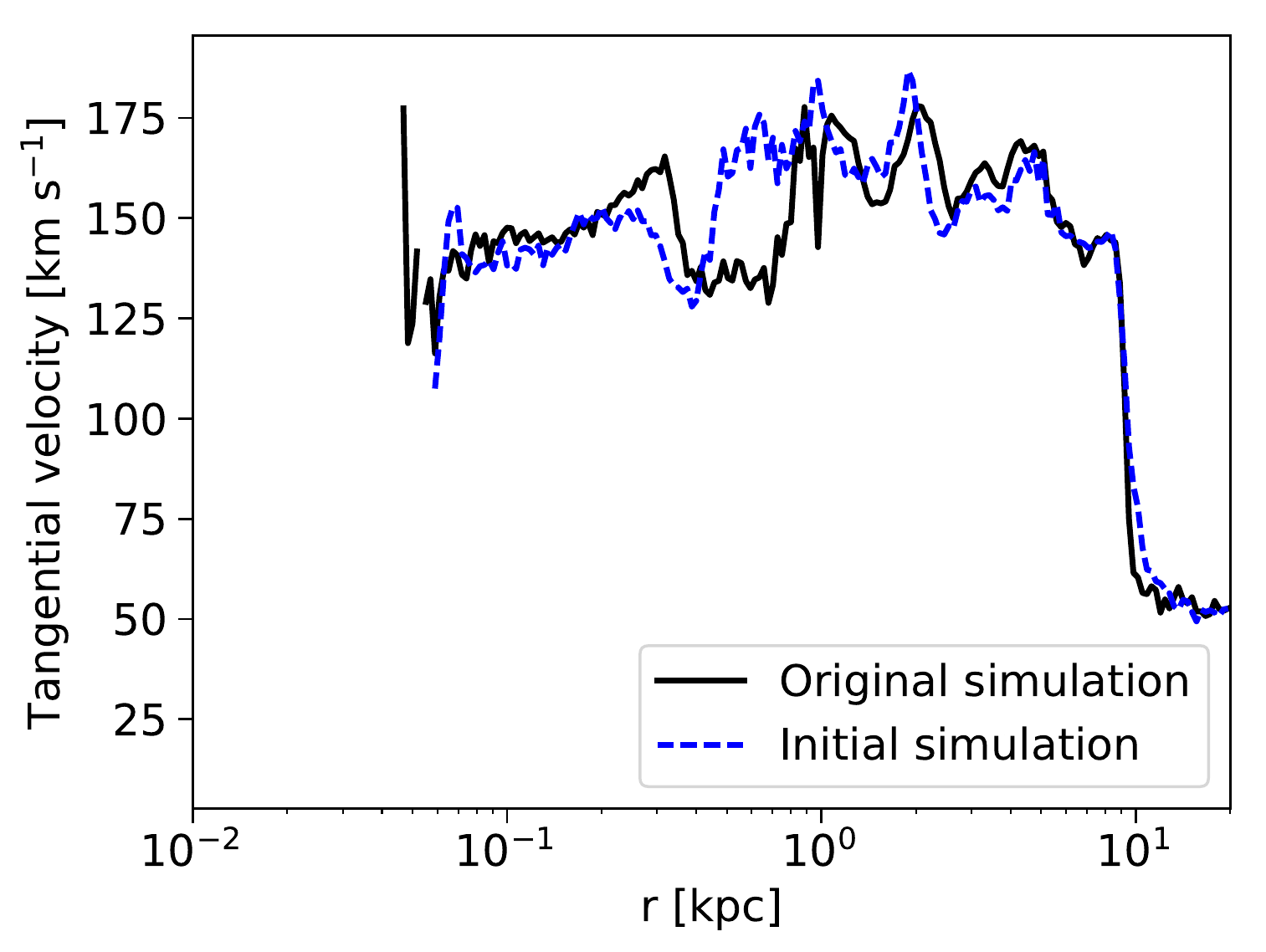}
 \includegraphics[width=\columnwidth]{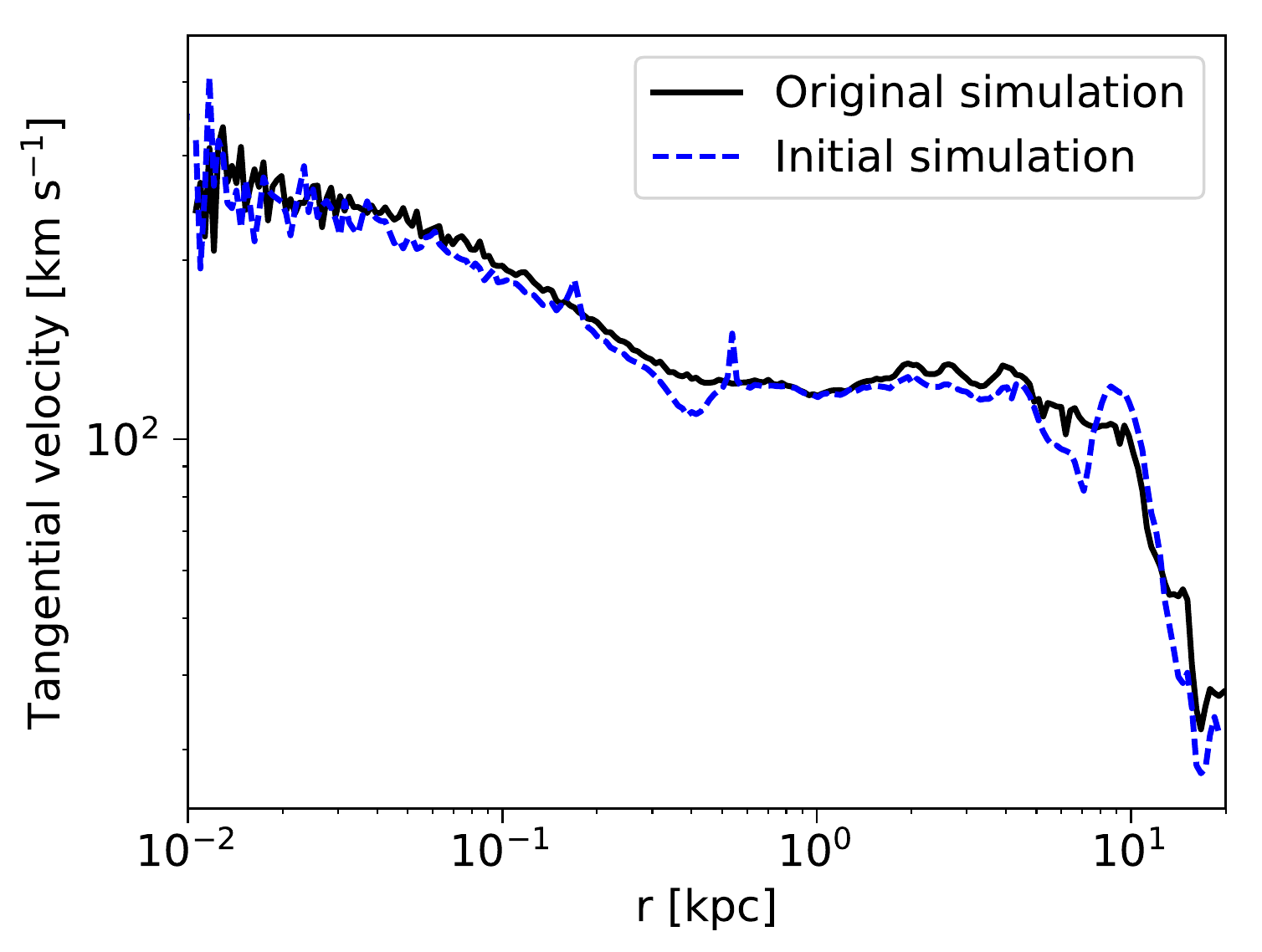}\\
 \includegraphics[width=\columnwidth]{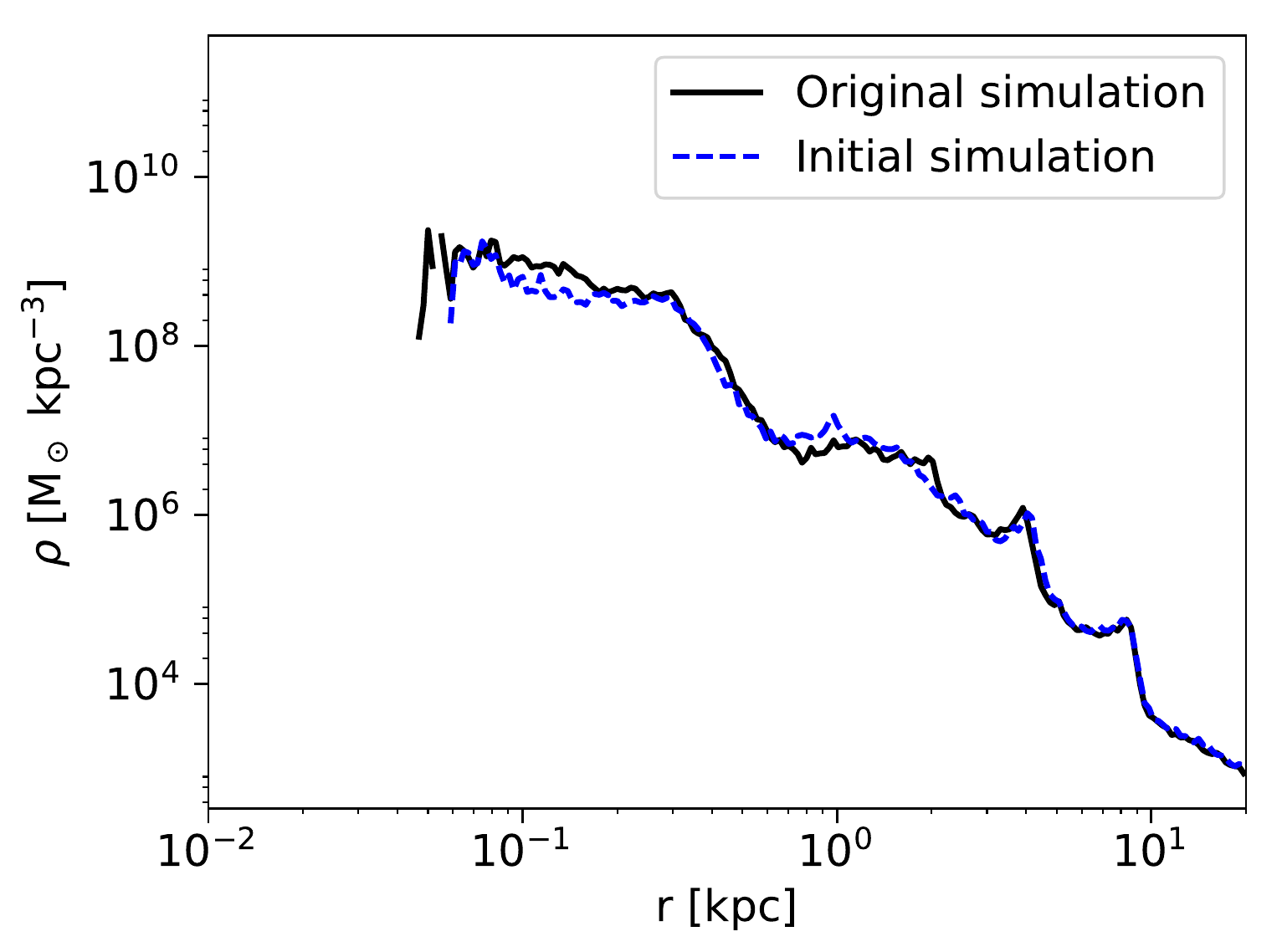}
 \includegraphics[width=\columnwidth]{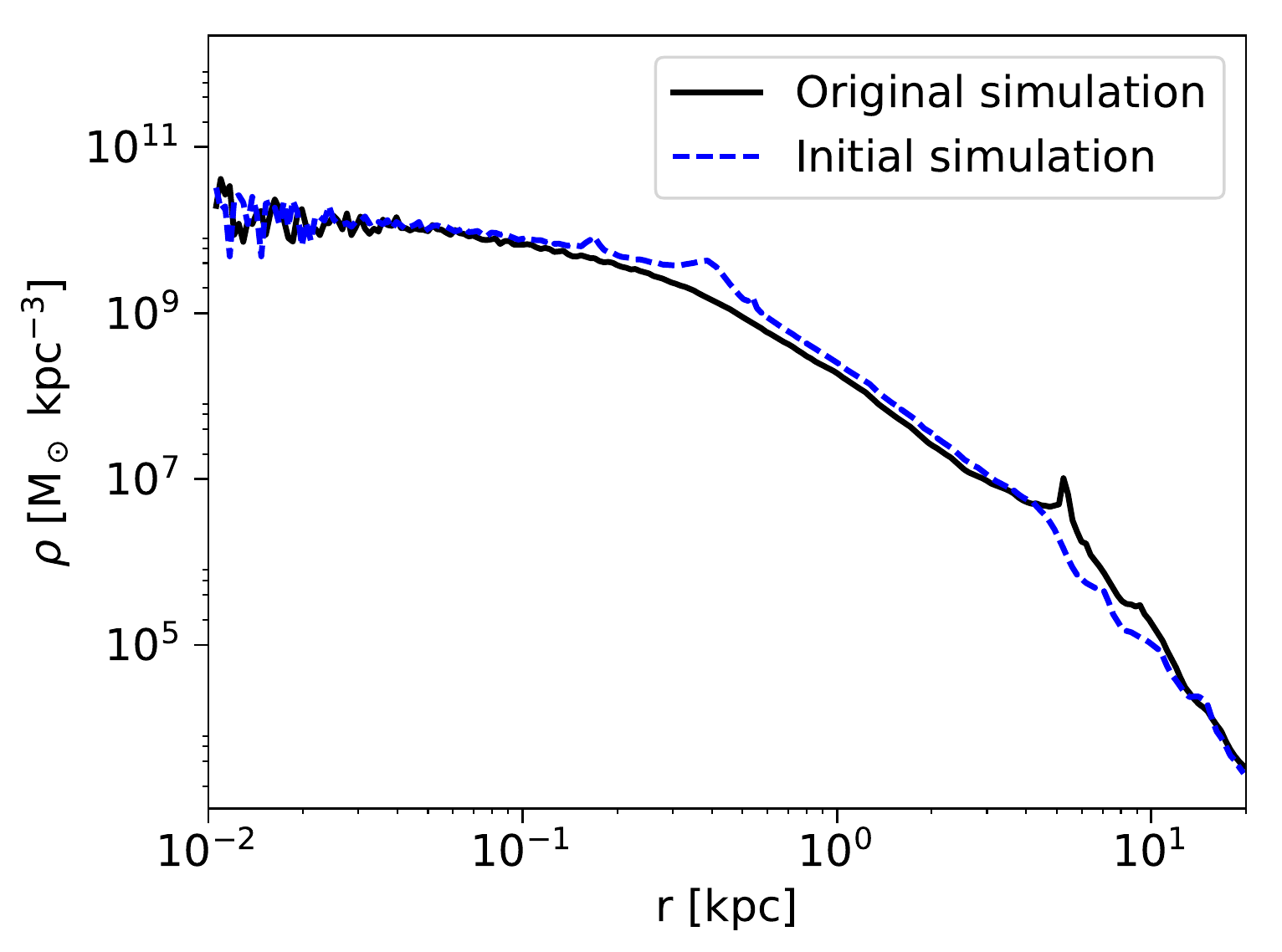}
 \caption{Radial velocity (top panels), tangential velocity (middle panels), and density (bottom panels) for gas (left-hand panels) and stars (right-hand panels) at $t=90\Myr$ in the original simulation (black, solid line) from \protect \cite{Capelo_15} and in our initial simulation (blue, dashed line), where we removed the outer particles of the remnant galaxy. The agreement in the inner 10 kpc is very good and we can safely trust our result for our zoomed simulations which are run for 30 Myr.}
 \label{AvQttys}
\end{figure*}

We present here details about the different tests we have performed to ensure that removing particles that are more than 20 kpc away from the new galactic centre does not affect the results. We have run a simulation that only differs from the original one from \cite{Capelo_15} by the removed particles, for a time much longer (90 Myr) than the 30 Myr for which we run our high-resolution simulations. We then compare different quantities, namely the distance between the two SMBHs as a function of time (Fig. \ref{DistPedroAndI}), the radial/tangential velocity and the gas and star density profiles at t=90 Myr, i.e. the last output (Fig. \ref{AvQttys}). Apart from the radial velocity in the outskirt of the galaxy, which differs because there is no more external pressure and because the outer material which was flowing in has been removed, the profiles between the initial and the original simulations are similar in the region relevant to the SMBH dynamics, confirming that removing the outer particles does not affect our results.

\section{Effects of zooming-in on the star formation}
\label{sec:EffectsOfZoomingInOnTheStarFormation}

\begin{table*}
 \begin{tabular}{lccccccp{7.5cm}}
  \hline
	Name&
	$\epsilon_\text{gas}$&
	$\epsilon_\text{star}$&
	$\epsilon_\text{BH}$&
	$\rho_{\text{crit}}$&
	$\tau_\text{SMBHB}$&
	$\tau_\text{res}$&
	Description\\
	
	&
	pc&
	pc&
	pc&
	100 a.m.u. cm$^{-3}$&
	Myr&	
	Myr
	\\
	
	\hline
	\hline
	
	R5 & $5$ & 2.5 & 1.25 &$10$&27 & 14 &X\\

	R5\_ISFT & $5$ & 2.5 & 1.25 & 1 &23 & 22 & Same resolution as R5 but keeping the initial SF threshold (ISFT) from the original run.\\	
	
	\hline

 \end{tabular}
 \caption{Simulations performed to study the effects of $\rho_\text{crit}$. We vary the density threshold for SF, $\rho_{\text{crit}}$. $\tau_\text{SMBHB}$ is the time at which the SMBHB is formed in our simulations and $\tau_\text{res}$ corresponds to the moment the distance between SMBHs is below $\epsilon_{\rm gas}$ for the first time. A description of the different simulations is also given.}
 \label{tab_simulation2}
\end{table*}

In Section \ref{Zoom_in}, we derived how to tune the parameter $\rho_\text{crit}$, the density threshold for SF, with the resolution of our simulation. In Section \ref{Explanation:DynamicalFriction}, we showed that in increased-resolution simulations the stellar density in the inner 20 pc is orders of magnitude higher than in the initial simulation. In principle, this increase could be caused by numerically-induced relaxation of the stellar particles in a very dense cusp or by an increased SF, although as the resolution increases, so does $\rho_\text{crit}$, compensating for the increased density as the softening is decreased.

To disentangle these effects, we ran a simulation (see Table \ref{tab_simulation2}) with similar properties to R5, but with $\rho_\text{crit}$ set to 100 a.m.u cm$^{-3}$, as in the initial simulation.

\begin{figure}
  \includegraphics[width=\columnwidth]{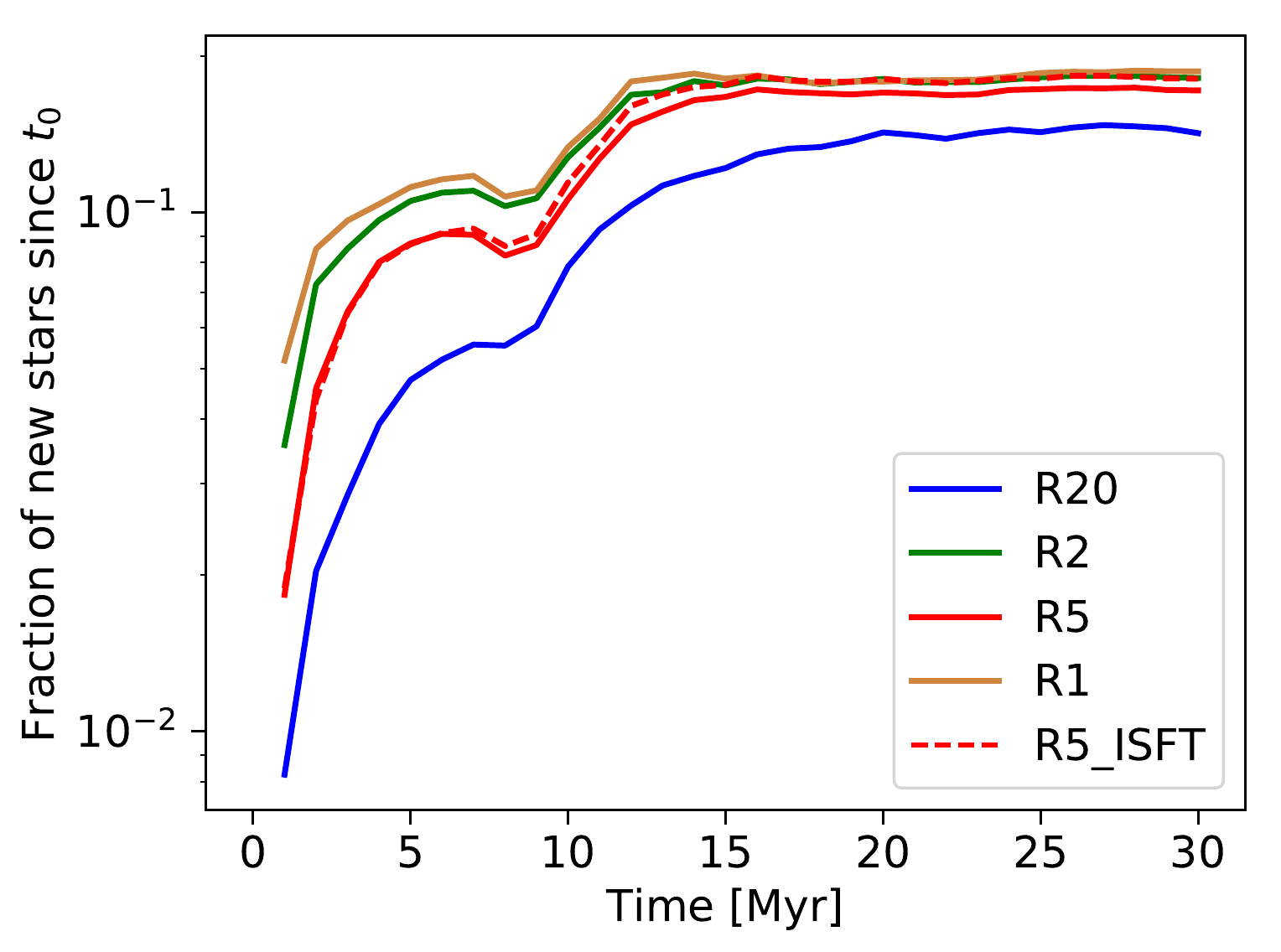}
 \caption{Mass fraction of new stars formed since $t_0$ for different simulations. We see that the precise value of $\rho_\text{crit}$ is not important (R5\_ISFT and R5 are similar) and that for the high-resolution simulations, the results converge.}
 \label{fig:FractionOfNewStars}
\end{figure}

In Fig. \ref{fig:FractionOfNewStars}, we show $M_{\star, \text{new}}/M_{\star}$, where $M_{\star, \text{new}}$ and $M_{\star}$ are, respectively, the mass of stars formed after $t_0$ and the total mass of stars, within 100 pc from BH2. Simulation R5\_ISFT is fairly similar to R5, meaning that, on 100-pc scales around the BHs, the precise value of $\rho_\text{crit}$ is not crucial. In fact, for the high-resolution cases, roughly 2/3 of the stars formed in the inner 100 pc actually form in the inner 10 pc, where, according to Fig. \ref{fig:DensityAllComponents}, the gas density is $\sim10^{12}\Mdot \kpc^{-3}\sim4\times 10^4$ a.m.u $\cm^{-3}$, which is larger than $\rho_\text{crit}$ in any case and explains why this parameter does not affect much the SF at this scale.

This actively star-forming region, in the inner 10 pc, which is below resolution in the initial simulation, leads to a difference in the SF, thus in the mass ratio between mass of new stars and total mass of stars, between the initial and the zoomed simulations. However, as for the density profiles (Fig. \ref{fig:DensityAllComponents}), we find a convergence in the zoomed simulations, with a similar behavior of the mass fraction of newly formed stars in all cases.

\bsp	
\label{lastpage}
\end{document}